\documentclass[a4paper,11pt]{article}
\pdfoutput=1 

\usepackage{jheppub} 
\usepackage{bm}
\usepackage{cancel}
\usepackage[T1]{fontenc} 
\usepackage{slashed}
\newcommand{\ri}{{\rm i}}

\title{\boldmath Roles of electric field/time-dependent Wilson line in toroidal compactification with or without magnetic fluxes}

\preprint{WU-HEP-24-03}

\author[a]{Hiroyuki Abe,}
\author[b]{Yusuke Yamada}

\affiliation[a]{Department of Physics, Waseda University, Tokyo 169-8555, Japan}
\affiliation[b]{Waseda Institute for Advanced Study, Waseda University, 1-21-1 Nishi Waseda, Shinjuku, Tokyo 169-0051, Japan}
\emailAdd{abe@waseda.jp}
\emailAdd{y-yamada@aoni.waseda.jp}

\abstract{We discuss the effects of electric fields along compact directions within (supersymmetric) gauge theory in $\mathbb{R}^{1,3}\times{\mathbb T}^2 (\times{\mathbb T}^2\times{\mathbb T}^2)$. The electric field along compact directions is equivalent to time-dependent homogeneous configuration of Wilson line moduli, which would be relevant to physics in the early universe. In particular, we consider models with and without background magnetic fluxes, which lead to completely different effects of the electric field due to the difference in the Kaluza-Klein (KK) level structure. We show that, in the case without magnetic fluxes, the deceleration of KK momenta may cause non-perturbative KK particle production dubbed as the KK Schwinger effect, whereas in the case with magnetic flux such KK particle production does not take place but flavor structure of low energy effective theory may be affected.}

\begin{document} 
\maketitle
\flushbottom

\section{Introduction}
Higher dimensional spacetime plays crucial roles in constructing realistic particle spectrum within string theory. For instance, flavor structure of particles in the standard model can be explained by the property of wave functions on compact extra spaces. One of the representative models is super Yang-Mills theory on tori with magnetic flux backgrounds~\cite{Cremades:2004wa,Abe:2008fi,Abe:2012fj,Abe:2012ya}, where the origin of the particle generation is the degenerate zero mode wave functions on tori and hierarchical Yukawa couplings can be realized by overlap integral of Gaussian like zero mode wave functions while the resultant gauge theory becomes chiral. 

Despite such fascinating properties of physics in compact extra spaces, there appear in four dimensional (4D) effective theory several problematic light fields associated with compact extra spaces, which are often called moduli fields. The most ubiquitous moduli fields originate from extra dimensional components of gauge or metric fields, which are regarded as scalar fields in 4D effective theory. As their mass terms are forbidden by gauge invariance of the tree level action, such moduli acquire masses only through subleading corrections via loops or non-perturbative effects. 
In the early universe within 4D effective theory, light scalar fields may move on their field spaces. For instance, moduli may play the role of an inflaton field~\cite{Arkani-Hamed:2003xts,Arkani-Hamed:2003wrq,Kaplan:2003aj,Avgoustidis:2006zp} but more generally moduli may be trapped at some point on their field space and eventually start to oscillate when their masses become comparable with the Hubble scale, which causes moduli problems.

What would happen if extra dimensional components of gauge fields (Wilson line moduli) have time-dependent field configuration? The time derivative of Wilson line moduli is understood as an ``electric field'' along the corresponding compact direction. One of the authors studied the quantum effects associated with such electric field within a simple 5D quantum electrodynamics (QED) model~\cite{Yamada:2024aca}.\footnote{See also \cite{Furuuchi:2015foh} for an earlier observation related to the KK Schwinger effect.} It turns out that the electric field along compactified space causes the acceleration/deceleration of a KK momentum and a KK mode at a given time may become a zero mode because of the deceleration. In such a case, the ``KK mode'' can be produced non-perturbatively, which we call the KK Schwinger effect as the higher-dimensional extension of the Sauter-Schwinger effect known in the context of strong field QED~\cite{Sauter:1931zz,Schwinger:1951nm}.\footnote{Note however that from 4D viewpoint, KK modes acquire time-dependent masses like the case of preheating in the early universe~\cite{Kofman:1997yn} or moduli trapping mechanism~\cite{Kofman:2004yc,Kikuchi:2023uqo}.} One of the most interesting properties of the KK Schwinger effect is that KK particles can be created even if the electric field energy is much below the compactification scale, namely the effect may take place within ``4D effective theory'' that is given by truncating ``would-be-heavy'' KK modes. Clearly, if KK modes are produced, such truncation is not justified and the ``4D effective theory'' breaks down. Such effects would be important for early universe models based on higher-dimensional spacetime theory as the ``break down'' implies possible signatures of compact dimensions in cosmology, and it has been known that particle production events during inflation leaves some observable imprints~\cite{Green:2009ds,Barnaby:2009dd,Barnaby:2010ke,Flauger:2016idt}. 

In this work, we study the role of time-dependent Wilson line moduli/electric fields within 6D (10D) supersymmetric models compactified on ${\mathbb T}^2$ (${\mathbb T}^2\times{\mathbb T}^2\times{\mathbb T}^2$). In particular, we consider the effect of magnetic fluxes, which are well motivated from the particle physics model building as mentioned. We show that the physical effects caused by the ``electric field'' are quite different in the presence or the absence of magnetic fluxes due to the different KK level structure. More specifically, as we will show below, the KK Schwinger effect described above does not take place within the magnetized models. Therefore, it turns out that time-dependent Wilson line can affect only zero modes within low energy effective theory. We show that within realistic particle physics models based on the magnetized model, the time-dependent Wilson lines affect flavor structure of the standard model particles such as time-dependent fermion masses.

As a technical tool, we use superspace formalism which is more suitable for stringy model building. Note however that, the effects we will describe below does not at all rely on supersymmetry and are applied to non-supersymmetric models in the same way. Nevertheless,  throughout this work, we will take time-dependent moduli background chosen by hand, which may be justified only when moduli are really light. Such an assumption will be a consequence of hidden (extended) supersymmetry which cancels the leading order quantum correction to the moduli potential.

The rest of the paper is organized as follows. In the next section, we show a general setup of the model we consider and discuss the purely electric field background case, namely the case without magnetic fluxes. The background gauge potential yields time-dependent superpotential couplings, which become time-dependent masses of component fields. We show a few examples of the KK Scwhinger effect. In Sec.~\ref{withmagneticflux}, we proceed to the case with magnetic fluxes. The gauge potential can be regarded as time-dependent Wilson lines, which enable us to find exact mode functions, and we show that the time-dependent couplings associated with the electric field do not cause the KK Schwinger effect unlike the pure electric field case. Nevertheless, it turns out that the time-dependent Wilson line changes the flavor structure of the standard model sector within 10D super Yang-Mills model on magnetized tori. We conclude in Sec.~\ref{conclusion}. We also give appendices for technical tools used in this work, which occupies half of this paper. In appendix~\ref{notation} we show the notation for the spinorial calculation, which is particularly used in appendix~\ref{appB} where the quantization of Dirac and complex scalar fields in time-dependent backgrounds is reviewed. These formulas can be applied to other models almost in the same way. In appendix~\ref{externalpotential}, we show an approach to deal with the electric field in the magnetized models, which is analogous to the pure electric background case but is practically less useful than that in Sec.~\ref{withmagneticflux}. Appendix~\ref{10Dreview} is devoted for the review of semi-realistic models based on 10D super Yang-Mills models.

\section{Setup and pure electric background case}\label{pureelectric}
We consider a supersymmetric U(1) gauge theory with a hypermultiplet in 6D compactified on $\mathbb{T}^2$. For simplicity, we assume the metric $ds^2=\eta_{\mu\nu}dx^\mu dx^\nu+(\pi R)^2((dy^4)^2+(dy^5)^2)=\eta^{\mu\nu}dx^\mu dx^\nu+(2\pi R)^2dz d\bar{z}$ where $\eta_{\mu\nu}={\rm diag}(-1,1,1,1)$, $z=\frac12 (y^4+\ri y^5)$ with the periodic boundary condition $z\sim z+1$ and $z\sim z+\ri$ or equivalently $y^4\sim y^4+2, \ y^5\sim y^5+2$. The action of a charged hypermultiplet $(Q,\tilde{Q})$ and a U(1) vector multiplet $(V,\phi)$ can be written in terms of 4D superspace as~\cite{Marcus:1983wb,Arkani-Hamed:2001vvu}
\begin{align}
    S=&\int d^6X\sqrt{-G}\Biggl[\frac{1}{4}\left\{\int d^2\theta W^\alpha W_\alpha+{\rm h.c.}\right\}+\int d^4\theta\frac{1}{(\pi R)^2}\left((\sqrt2\bar\partial V-\bar\phi)(\sqrt2 \partial V-\phi)-\partial V\bar\partial V\right)\nonumber\\
    &+\int d^4\theta(\tilde{Q} e^{qgV}\bar{\tilde{Q}}+\bar{Q}e^{-qgV}Q)+\left\{\int d^2\theta \frac{1}{ \pi R}\tilde{Q}\left(\partial-\frac{qg}{\sqrt2}\phi\right)Q+{\rm h.c.}\right\}\Biggr],\label{6DS}
\end{align}
where $V$, $\phi$, $Q$ and $\tilde{Q}$ are, respectively, a vector and three chiral superfields of the 4D supersymmetry,  among those the lowest component of $\phi$ carries is a combination of the gauge field of the compact directions $\phi|_{\theta=\bar\theta=0}=\frac{1}{\sqrt2}(A_5+\ri A_4)\equiv A_z$ and we have used the complexified derivative $\partial\equiv\partial_4-\ri \partial_5(=\partial_z)$.\footnote{\label{3} Note that as the superfield containing 6D field, $V$ has mass dimension $1$ and  $g$ does $-1$ and therefore the exponent is dimensionless. Note also that in our convention, the coordinates of the compact spaces are dimensionless.} Here, the volume element $\sqrt{-G}$ in this case is $\sqrt{-G}=2(\pi R)^2$ in the complex coordinates and $q$ is a charge of the hypermultiplet.

We introduce a (spurion) background field for the gauge potential superfield along the torus $\phi$:
\begin{align}
    \langle\phi(x,\theta) \rangle=\phi_B(t,z,\bar{z})+\ri\theta\sigma^0\bar\theta \dot{\phi}_B-\frac{1}{4}\theta^2\bar\theta^2 \ddot{\phi}_B,
\end{align}
and then the superfield $\phi$ can formally be written as $\phi(x,\theta)= \delta\phi(x,\theta)+\langle\phi(x,\theta)\rangle$ where $\delta\phi(x,\theta)$ denotes the fluctuation around the background.
In particular, we consider an electro-magnetic background
\begin{align}
    \phi_B(t,z,\bar{z})=\frac{\pi N}{\sqrt2 qg}\bar{z}+\zeta(t),
\end{align}
where $N\in\mathbb Z$ denotes a quantized magnetic flux. The homogeneous zero mode $\zeta(t)$ leads to the electric field as
\begin{align}
   E_z\equiv \langle F_{0z}\rangle=\partial_t A_z(t)=\dot{\zeta}(t),
\end{align}
but one may also regard it as a Wilson line modulus field since we have\footnote{As the Abelian gauge field does not feel magnetic flux, their mode function becomes trivial even when magnetic fluxes are turned on.}
\begin{align}
   \theta(t) =&\oint_{C_1} dy^5 A_5(t,z)+\ri\oint_{C_2} dy^4 A_4(t,z)\nonumber\\
   =&\oint_{C_1} dy^5 A_5^{(0)}(t,z)+\ri\oint_{C_2} dy^4 A^{(0)}_4(t,z)\nonumber\\
   =&\sqrt2\zeta(t),
\end{align}
where $C_{1,2}$ are curves along ${\rm Re}z\to{\rm Re}z+1$ and ${\rm Im}z\to{\rm Im}z+1$, respectively and we have used the fact that only a zero mode $A_z^{(0)}(x)$ contribute to the integration. Thus we identify the time-dependent Wilson line with the electric field along a compact direction.

 We notice that since there is no coupling between $\phi$ and the hypermultiplet sector in the K\"ahler potential, only the lowest component of the spurion couples to the field in the hypermultiplet. Moreover, the K\"ahler potential terms that contain $\phi$ can be rewritten as
\begin{align}
    K\supset \frac{1}{(\pi R)^2}\int d^4\theta\left[\partial V\bar{\partial}V+\sqrt{2}(\bar{\partial}\phi+\partial\bar{\phi})V+\phi\bar{\phi}\right]
\end{align}
where we have performed integration by part with respect to $\partial$ and $\bar\partial$ acting on $V$. Thus, the Wilson line modulus $\zeta(t)$ contributes only to the last term, which yields the kinetic energy of the background, namely the electric field energy. Thus, the only nontrivial part of the time-dependent part appears as an effective mass term
\begin{align}
    -\frac{qg}{\sqrt2\pi R}\int d^2\theta \zeta(t)\tilde{Q}Q+{\rm h.c.}.
\end{align}

In the following sections, we perform KK expansion within superspace both for the case with and without magnetic fluxes. Such techniques have been used in the literature~\cite{Buchmuller:2016gib,Buchmuller:2018eog,Buchmuller:2019zhz,Buchmuller:2020nnl}, which yields manifestly supersymmetric description of KK expansion within our framework.

\subsection{Pure electric field model}
Let us first discuss the case without magnetic flux $N=0$. In this case the KK expansion of superfields is\footnote{For simplicity, we are concerned with the square torus and one notices that the following expansion is just Fourier expansion by plane waves.}
\begin{align}
    \Phi(x,\theta,\bar\theta, z,\bar{z})=&\frac{1}{2\pi R}\sum_{n_R,n_I\in \mathbb{Z}}\Phi_{n_R,n_I}(x,\theta,\bar\theta)\exp\left(2\ri\pi n_R{\rm Re}z+2\ri\pi n_I{\rm Im}z\right)\nonumber\\
    =&\frac{1}{2\pi R}\sum_{n_R,n_I\in \mathbb{Z}}\Phi_{n_R,n_I}(x,\theta,\bar\theta)\exp\left(\pi (\ri n_R+n_I)z+\pi(\ri n_R-n_I)\bar{z}\right)\label{PhiKK}
\end{align}
and
\begin{align}
    \bar{\Phi}(x,\theta,\bar\theta, z,\bar{z})=&\frac{1}{2\pi R}\sum_{n_R,n_I\in \mathbb{Z}}\bar{\Phi}_{n_R,n_I}(x,\theta,\bar\theta)\exp\left(-2\ri\pi n_R{\rm Re}z-2\ri\pi n_I{\rm Im}z\right)\nonumber\\
    =&\frac{1}{2\pi R}\sum_{n_R,n_I\in \mathbb{Z}}\bar{\Phi}_{n_R,n_I}(x,\theta,\bar\theta)\exp\left(\pi(-\ri n_R-n_I)z+\pi (-\ri n_R+n_I)\bar{z}\right)\label{bPhiKK}
\end{align}
for all superfields. Note that $V_{n_R,n_I}=\bar{V}_{-n_R,-n_I}$ for a real superfield $V$. With the above KK expansion, one finds the following set of 4D action: The gauge kinetic term becomes
\begin{align}
    \int d^6X\sqrt{-G}\frac{1}{4}\left\{\int d^2\theta W^\alpha W_\alpha+{\rm h.c.}\right\}=\frac{1}{4}\int d^4x \sum_{n_R,n_I\in \mathbb{Z}}\left\{\int d^2\theta \tilde{W}^{\alpha}_{n_R,n_I}W_{n_R,n_I\alpha}+{\rm h.c.}\right\}
\end{align}
where $W^{\alpha}_{n_R,n_I}$ and $\tilde{W}^{\alpha}_{n_R,n_I}$ are the chiral field strength of $V_{n_R,n_I}$ and $\bar{V}_{n_R,n_I}$, respectively.
Similarly, integration over compact spaces reads
\begin{align}
&  \frac{1}{(\pi R)^2}\int d^6X\sqrt{-G}\int d^4\theta\left[\partial V\bar{\partial}V+\sqrt{2}(\bar{\partial}\phi+\partial\bar{\phi})V+\phi\bar{\phi}\right]\nonumber\\
=&\int d^4xd^4\theta\sum_{n_R,n_I\in\mathbb{Z}}\Biggl[|M_{n_R,n_I}|^2\bar{V}_{n_R,n_I}V_{n_R,n_I} -\frac{\sqrt{2}}{\pi R}\{\bar{M}_{n_R,n_I}\delta\phi_{n_R,n_I}\bar{V}_{n_R,n_I}+M_{n_R,n_I}\overline{\delta\phi}_{n_R,n_I}V_{n_R,n_I}\}\nonumber\\
&\qquad +\frac{1}{(\pi R)^2}\left|\delta\phi_{n_R,n_I}+2\pi R\phi_B\delta_{n_R,0}\delta_{n_I,0}\right|^2\Biggr],
\end{align}
where $M_{n_R,n_I}\equiv (n_I+\ri n_R)/R$, $(n_R,n_I)\neq(0,0)$ for $\phi_{n_R,n_I}$ and its conjugate, and $\delta_{n,m}$ denotes Kronecker delta. Canonical normalization of $\phi_{n_R,n_I}$ is achieved by the replacement $\delta\phi_{n_R,n_I}\to\pi R\delta\phi_{n_R,n_I}$ such that mass dimension of $\delta\phi_{n_R,n_I}$ becomes $1$.\footnote{This is necessary since KK expansion \eqref{PhiKK} reads the mass dimension of $\delta\phi_{n_R,n_I}$ to be $0$ since the original 6D superfield $\phi$ has dimension 1. (Regarding $V$, see footnote~\ref{3}).} 
Shifting vector multiplets $V_{n_R,n_I}\to V_{n_R,n_I}+\frac{1}{\sqrt{2}M_{n_R,n_I}}\delta\phi_{n_R,n_I}-\frac{1}{\sqrt{2}\bar{M}_{n_R,n_I}}\overline{\delta\phi}_{n_R,n_I}$ for $(n_R,n_I)\neq (0,0)$ simplifies the above terms and compactly summarizes the 4D action as
\begin{align}
\int d^4x\Biggl[&\sum_{n_R,n_I\in\mathbb{Z}}\left\{\int d^2\theta\frac14\tilde{W}^{\alpha}_{n_R,n_I}W_{n_R,n_I\alpha}+{\rm h.c.}\right\}\nonumber\\
&+\int d^4\theta\Biggl\{\sum_{ n_R,n_I\in\mathbb{Z}\setminus (0,0)}|M_{n_R,n_I}|^2|V_{n_R,n_I}|^2+|\delta\phi_{0,0}+2\phi_B|^2\Biggr\}\Biggr],
\end{align}
and note that the zero mode cannot be ``eaten'' by the KK modes of vector superfields whereas other KK modes of $\phi$ are eaten by the KK vector superfields.

We proceed to the hypermultiplet action.
Notice that the shift of vector multiplet can be regarded as gauge transformation and therefore it should not change the hypermultiplet action. Then, taking the Wess-Zumino gauge in which $V^3=0$, we obtain
\begin{align}
   &\int d^6X\sqrt{G}\Biggl[\int d^4\theta\left((1+q^2g^2V^2/2)(\tilde{Q}\bar{\tilde{Q}}+\bar{Q}Q)+gq V(\tilde{Q}\bar{\tilde{Q}}-\bar{Q}Q)\right)\nonumber\\
    &\qquad+\left\{\int d^2\theta \frac{1}{ \pi R}\tilde{Q}\left(\partial-\frac{q g\phi_B}{\sqrt2}-\frac{qg\pi R}{\sqrt2}\phi\right)Q+{\rm h.c.}\right\}\Biggr]\nonumber\\
    =&\int d^4x\sum_{n_R,n_I\in \mathbb{Z}}\Biggl[\int d^4\theta\left(\bar{\tilde{Q}}_{n_R,n_I}e^{qg_{\rm 4D}V_{0,0}}\tilde{Q}_{n_R,n_I}+\bar{Q}_{n_R,n_I}e^{-qg_{\rm 4D}V_{0,0}}Q_{n_R,n_I}\right)\nonumber\\
  &  +\left\{\int d^2\theta\left(M_{n_R,n_I}-\sqrt2 qg_{4D}\phi_B\right)\tilde{Q}_{-n_R,-n_I}Q_{n_R,n_I}+{\rm h.c.}\right \}\Biggr]+\cdots,
\end{align}
where we have defined $g_{\rm 4D}\equiv g/(2\pi R)$ and the ellipses denote the terms containing interactions between hypermultiplets and KK vector multiplets, which we will not focus on. We have also omitted the cubic interaction between hypermultiplets and the fluctuation of a Wilson line modulus $\delta\phi_{0,0}$. Noting that there appear no quadratic terms associated with D-term tadpole in the absence of magnetic fluxes, we find the quadratic action
\begin{align}
    S_2=\int d^4x\sum_{n_R,n_I\in \mathbb{Z}}\Biggl[\int d^4\theta\left(|\tilde{Q}_{n_R,n_I}|^2+|Q_{n_R,n_I}|^2\right) +\left\{\int d^2\theta M_{n_R,n_I}^{\rm eff}\tilde{Q}_{-n_R,-n_I}Q_{n_R,n_I}+{\rm h.c.}\right \}\Biggr],
\end{align}
where $M_{n_R,n_I}^{\rm eff}=M_{n_R,n_I}-\sqrt2qg_{4D}\phi_B$. Interestingly, at the quadratic order, the mass term of the hypermultiplet is supersymmetric, namely, the mass of bosons and fermions in the hypermultiplet are degenerate. This does not mean supersymmetry is unbroken since the background gauge potential field $\phi$ is time-dependent and supersymmetry transformation of a fermionic component of $\phi$ is non-vanishing.\footnote{Indeed, it is possible to spontaneously break supersymmetry by time-translation breaking~\cite{Yamada:2021kxv}.}

\subsection{KK Schwinger effect}
We discuss how KK particles can be produced from vacuum non-perturbatively. Let us have a closer look at the effective KK mass term
\begin{align}
    M_{n_R,n_I}^{\rm eff}=M_{n_R,n_I}-\sqrt2qg_{4D}\phi_B=\frac{1}{R}(n_I+\ri n_R)-qg_{\rm 4D}(\langle A_6+\ri A_5\rangle),\label{effKKmass}
\end{align}
where $A_{5,6}$ are original 6D gauge potential.\footnote{Let us note about dimension of each: $g_{4D}$ is dimensionless while $A_{5,6}$ has mass dimension $1$. Notice also that the original 6D gauge coupling has mass dimension $-1$.} Recall that the mass term originates from the momenta along compact direction, and the time-dependent gauge potential leads to electric forces along torus direction, which is the reason why $\tilde{Q}_{n_R,n_I}$ has the effective mass $M_{-n_R,-n_I}^{\rm eff}$ as it has the charge opposite to $Q_{n_R,n_I}$.

The time-dependent mass generally cause non-perturbative pair production of hypermultiplet particles via the KK Schwinger effect, which is analogous to the Schwinger effect in 4D~\cite{Sauter:1931zz,Schwinger:1951nm}.\footnote{The real time evaluation of particle number in a compactified space is also discussed in \cite{Qiu:2020gbl}. }  If the modulus $\zeta$ varies over a scale larger than the KK scale $M_{\rm KK}\sim 1/R$, some of ``KK particles'' become massless at certain time, which can be understood as deceleration of a KK momentum by the electric field. In order to clarify it, we perform $\theta$-integration and obtain the quadratic action of fluctuations
\begin{align}
    S_2=\sum_{{\bm n}\in\mathbb{Z}^2}\int d^4x \Biggl[&-|\partial_\mu Q_{\bm n}|^2-|\partial_\mu \tilde{Q}_{\bm n}|^2-\ri\bar{\psi}_{\bm n}\bar{\sigma}^\mu\partial_\mu\psi_{\bm n}-\ri\bar{\tilde{\psi}}_{\bm n}\bar{\sigma}^\mu\partial_\mu\tilde{\psi}_{\bm n}\nonumber\\
    &-(M^{\rm eff}_{\bm n}\tilde{\psi}_{\bm n}\psi_{\bm n}+{\rm h.c.})-|M_{\bm n}^{\rm eff}|^2(|Q_{\bm n}|^2+|\tilde{Q}_{\bm n}|^2)\Biggr],
\end{align}
where we have performed the integration of F- and D-terms. Here we have used the component expansion of superfields,
\begin{align}
    Q_{\bm n}(x,\theta,\bar\theta)=Q_{\bm n}(x)+\sqrt{2}\theta\psi_{\bm n}(x)+\theta\theta F_{\bm n}(x)+\cdots
\end{align}
where $\bm n=(n_R,n_I)$, the ellipses denote terms that vanish in the chiral coordinates $y^\mu=x^\mu+\ri\theta\sigma^\mu\bar\theta$ and $\tilde{Q}_{\bm n}$ is expanded similarly.
The Klein-Gordon and the Dirac equation are
\begin{align}
   & -\partial^2Q_{\bm n}+\left|M_{\bm n}^{\rm eff}\right|^2Q_{\bm n}=0,\\
   &-\ri\bar{\sigma}^\mu\partial_\mu\psi_{\bm n}-\bar{M}_{\bm n}^{\rm eff}\bar{\tilde{\psi}}_{\bm n}=0,\\
   &-\ri\sigma^\mu\partial_\mu\bar{\tilde\psi}_{\bm n}-M_{\bm n}^{\rm eff}\psi_{\bm n}=0.
\end{align}

Let us consider the quantization of the fields having time-dependent KK masses. The quantization procedure of both Weyl spinors and complex scalars having a time-dependent (KK) mass is summarized in appendix~\ref{appB}. Using those results to our models,  we expand each KK mode as
\begin{align}
    \hat{\psi}^\alpha_{\bm n}(t,\bm x)=&\sum_{h=\pm}\int \frac{d^3\bm k}{(2\pi)^\frac32}\left[\hat{c}_{\bm k,\bm n,h}\chi_{k,\bm n,h}(t)e^{\ri\bm k\cdot\bm x}\xi_{h}^\alpha(\hat{\bm k})+\hat{d}^\dagger_{\bm k,\bm n,h}\bar{\eta}_{k,\bm n,h}(t)e^{-\ri{\bm k}\cdot\bm x}\xi^{\dagger}_{h\dot\beta}(\hat{\bm k})\bar\sigma_0^{\dot\beta\alpha}\right],\label{psn}\\
    \hat{\tilde\psi}^\alpha_{\bm n}(t,\bm x)=&\sum_{h=\pm}\int \frac{d^3\bm k}{(2\pi)^\frac32}\left[\hat{d}_{\bm k,\bm n,h}\chi_{k,\bm n,h}(t)e^{\ri\bm k\cdot\bm x}\xi_h^\alpha(\hat{\bm k})+\hat{c}^\dagger_{\bm k,\bm n,h}\bar{\eta}_{k,\bm n,h}(t)e^{-\ri{\bm k}\cdot\bm x}\xi^{\dagger}_{h\dot\beta}(\hat{\bm k})\bar\sigma_0^{\dot\beta\alpha}\right]\label{pstn},\\
    \hat{Q}_{\bm n}(t,\bm x)=&\int\frac{d^3\bm k}{(2\pi)^{\frac32}}\left(\hat{a}_{\bm k,\bm n}f_{k,\bm n}(t)e^{\ri\bm k\cdot\bm x}+\hat{b}^\dagger_{\bm k,\bm n}\bar{f}_{k,\bm n}(t)e^{- \ri\bm k\cdot\bm x}\right),\label{qn}
\end{align}
where $\hat{a}_{\bm k,\bm n},\hat{b}_{\bm k,\bm n},\hat{c}_{\bm k,\bm n,h}, \hat{d}_{\bm k,\bm n,h}$ are the annihilation operator and their conjugates are corresponding creation operators satisfying $[\hat{a}_{\bm k,\bm n},\hat{a}^\dagger_{\bm k',\bm n'}]=\delta^3(\bm k-\bm k')\delta_{\bm n \bm n'}$, $[\hat{b}_{\bm k,\bm n},\hat{b}^\dagger_{\bm k',\bm n'}]=\delta^3(\bm k-\bm k')\delta_{\bm n \bm n'}$, $\{\hat{c}_{\bm k,\bm n},\hat{c}^\dagger_{\bm k',\bm n'}\}=\delta^3(\bm k-\bm k')\delta_{\bm n \bm n'}$, $\{\hat{d}_{\bm k,\bm n},\hat{d}^\dagger_{\bm k',\bm n'}\}=\delta^3(\bm k-\bm k')\delta_{\bm n \bm n'}$ and the label $h$ denotes the helicity of Weyl spinors with the heilicity eigenspinors $\xi_{h}^\alpha(\hat{\bm k}),\xi^{\dagger}_{h\dot\beta}(\hat{\bm k})$ which depend only on $\hat{\bm k}={\bm k}/|\bm k|$ (for $|\bm k|\neq 0$).  The mode functions $f_{k,\bm n}(t), \chi_{k,\bm n,h}(t),\eta_{k,\bm n,h}(t)$ satisfy, respectively,
\begin{align}
&\ddot{f}_{k,\bm n}+(k^2+|M_{\bm n}^{\rm eff}(t)|^2)f_{k,\bm n}=0,\\
 &  \ri\frac{d}{dt}\left(\begin{array}{c}\chi_{k,\bm n,h}(t)\\ \eta_{k,\bm n,h}(t)\end{array}\right)=\left(\begin{array}{cc}
        -kh & -\bar{M}_{\bm n}^{\rm eff}(t) \\
       -M_{\bm n}^{\rm eff}(t)  & kh
    \end{array}\right)\left(\begin{array}{c}\chi_{k,\bm n,h}(t)\\ \eta_{k,\bm n,h}(t)\end{array}\right).
\end{align}
These mode equations can be formally solved as
\begin{align}
   &\left(\begin{array}{c}\chi_{k,\bm n,h}(t)\\ \eta_{k,\bm n,h}(t)\end{array}\right)=\alpha_{k,\bm n,h}(t)e^{-\ri\int^t dt'\omega_{k,\bm n}(t')}{\bm v}_{k,\bm n,h}^+(t)+\beta_{k,\bm n,h}(t)e^{+\ri\int^t dt'\omega_{k,\bm n}(t')}{\bm v}_{k,\bm n,h}^-(t),\label{KKFadexpand}\\
&f_{k,\bm n}(t)=\frac{1}{\sqrt{2\omega_{k,\bm n}(t)}}\left[\gamma_{k,\bm n}(t)e^{-\ri\int^tdt'\omega_{k,\bm n}(t')}+\delta_{k,\bm n}(t)e^{\ri\int^tdt'\omega_{k,\bm n}(t')}\right],\label{KKBadexpand}
\end{align}
where we have parametrized $M_{\bm n}^{\rm eff}(t)=\mu_{\bm n}(t)e^{\ri\rho_{\bm n}(t)}$ where  $\mu_{\bm n}(t)=|M_{\bm n}^{\rm eff}(t)|$ and $\rho_{\bm n}(t)={\rm Arg}M_{\bm n}^{\rm eff}(t)$ and defined
\begin{align}
&\omega_{k,\bm n}(t)\equiv\sqrt{k^2+\mu_{\bm n}^2(t)},\\
&\cos\theta_{k,\bm n,h}(t)\equiv-\frac{kh}{\omega_{k,\bm n}(t)},\quad \sin\theta_{k,\bm n,h}(t)\equiv-\frac{\mu_{\bm n}(t)}{\omega_{k,\bm n}(t)},\\
&    {\bm v}_{k,\bm n,h}^+(t)\equiv \left(\begin{array}{c} e^{-\ri\rho_{\bm n}(t)}\cos\frac{1}{2}\theta_{k,\bm n,h}(t)\\ \sin\frac{1}{2}\theta_{k,\bm n,h}(t)
    \end{array}\right),\qquad {\bm v}_{k,\bm n,h}^-(t)\equiv\left(\begin{array}{c} -e^{-\ri\rho_{\bm n}(t)}\sin\frac{1}{2}\theta_{k,\bm n,h}(t)\\ \cos\frac{1}{2}\theta_{k,\bm n,h}(t)
    \end{array}\right).
\end{align}
The auxiliary functions $\alpha_{k,\bm n,h}(t)$, $\beta_{k,\bm n,h}(t)$, $\gamma_{k,\bm n}(t)$, $\delta_{k,\bm n}(t)$ satisfy the normalization conditions
\begin{align}
 |\alpha_{k,\bm n,h}(t)|^2+|\beta_{k,\bm n,h}(t)|^2=1, \qquad |\gamma_{k,\bm n}(t)|^2-|\delta_{k,\bm n}(t)|^2=1
\end{align}
following from the canonical (anti)commutation relations of the field operators and that of the creation and annihilation operators.

The formally introduced creation and annihilation operators such as $\hat{a}_{\bm k,\bm n,h}$ can be interpreted as that of past adiabatic vacuum $|0\rangle_{\rm in}$ by imposing the asymptotic conditions to the auxiliary functions as
\begin{align}
    \alpha_{k,\bm n,h}(t)\to 1,\quad \beta_{k,\bm n,h}(t)\to 0,\quad \gamma_{k,\bm n}(t)\to 1,\quad \delta_{k,\bm n}(t)\to 0
\end{align}
for $t\to -\infty$. With these conditions, one can confirm that the creation and annihilation operators become coefficients of either positive or negative frequency adiabatic mode functions at the asymprotic past. 

The time-dependent auxiliary functions are understood as time-dependent Bogoliubov coefficients as explained in detail in appendix~\ref{appB}, and the particle number density produced from the past adiabatic vacuum can be quantitatively evaluated as
\begin{align}
   & {}_{\rm in}\langle 0|\hat{N}^{\psi}_{\bm k,\bm n,h}(t)|0\rangle_{\rm in}={}_{\rm in}\langle 0|\hat{N}^{\tilde{\psi}}_{\bm k,\bm n,h}(t)|0\rangle_{\rm in}=|\beta_{k,\bm n,h}(t)|^2,\\
   & {}_{\rm in}\langle 0|\hat{N}^{Q}_{\bm k,\bm n}(t)|0\rangle_{\rm in}=|\delta_{k,\bm n}(t)|^2,
\end{align}
where $\hat{N}^{\psi,\tilde\psi}_{\bm k,\bm n,h}(t)$ and $\hat{N}^{Q}_{\bm k,\bm n}(t)$ are number density operators of each field and $|0\rangle_{\rm in}$ is the past adiabatic vacuum annihilated by any of $\hat{a}_{\bm k,\bm n},\hat{b}_{\bm k,\bm n},\hat{c}_{\bm k,\bm n,h}, \hat{d}_{\bm k,\bm n,h}$. The time-dependent Bogoliubov coefficients obey
\begin{align}
&\frac{d}{dt}\left(\begin{array}{c}\tilde{\alpha}_{k,\bm n,h}\\ \tilde{\beta}_{k,\bm n,h}\end{array}\right)=\left(\begin{array}{cc}-\ri\omega_{k,\bm n}+\frac{\ri}{2}\dot{\phi}_{\bm n}(1+\cos\theta_{k,\bm n,h})&\ \frac12\left(\dot\theta_{k,\bm n,h}-\ri\dot\phi_{\bm n}\sin\theta_{k,\bm n,h}\right)\\
-\frac12\left(\dot{\theta}_{k,\bm n,h}+\ri\dot{\phi}_{\bm n}\sin\theta_{k,\bm n,h}\right)&\ri\omega_{k,\bm n}+\frac{\ri}{2}\dot\phi_{\bm n}(1-\cos\theta_{k,\bm n,h})\end{array}\right)\left(\begin{array}{c}\tilde{\alpha}_{k,\bm n,h}\\ \tilde{\beta}_{k,\bm n,h}\end{array}\right),\\
&\frac{d}{dt}\left(\begin{array}{c}\tilde{\gamma}_{k,\bm n}\\ \tilde{\delta}_{k,\bm n}
    \end{array}\right)=\left(\begin{array}{cc}\ri\omega_{k,\bm n}&\frac{\dot{\omega}_{k,\bm n}}{2\omega_{k,\bm n}}\\
    \frac{\dot{\omega}_{k,\bm n}}{2\omega_{k,\bm n}}&-\ri\omega_{k,\bm n}\end{array}\right)\left(\begin{array}{c}\tilde{\gamma}_{k,\bm n}\\ \tilde{\delta}_{k,\bm n}
    \end{array}\right),
\end{align}
where we have introduced $\tilde{\alpha}_{k,\bm n,h}\equiv \alpha_{k,\bm n,h}e^{-\ri\int^tdt'\omega_{k,\bm n}(t')}$, $\tilde{\beta}_{k,\bm n,h}\equiv \beta_{k,\bm n,h}e^{\ri\int^tdt'\omega_{k,\bm n}(t')}$, $\tilde{\gamma}_{k,\bm n}\equiv \gamma_{k,\bm n}e^{-\ri\int^tdt'\omega_{k,\bm n}(t')}$, and $\tilde{\delta}_{k,\bm n}\equiv \delta_{k,\bm n}e^{\ri\int^tdt'\omega_{k,\bm n}(t')}$. It is generally hard to find analytical solutions to the above differential equations. Nevertheless, by analyzing the Stokes phenomena of differential equations, it is possible to know the amount of particle production as well as the ``time'' of particle production~\cite{Dumlu:2010ua,Dabrowski:2014ica,Dabrowski:2016tsx,Li:2019ves,Enomoto:2020xlf,Taya:2020dco,Hashiba:2021npn,Enomoto:2021hfv,Yamada:2021kqw,Enomoto:2022nuj,Enomoto:2022mti,Hashiba:2022bzi}. We here do not discuss Stokes phenomena and just solve the above equations numerically for some choices of $\zeta(t)$ below.

\begin{figure}[htbp]
    \centering
\includegraphics[keepaspectratio, scale=0.6]{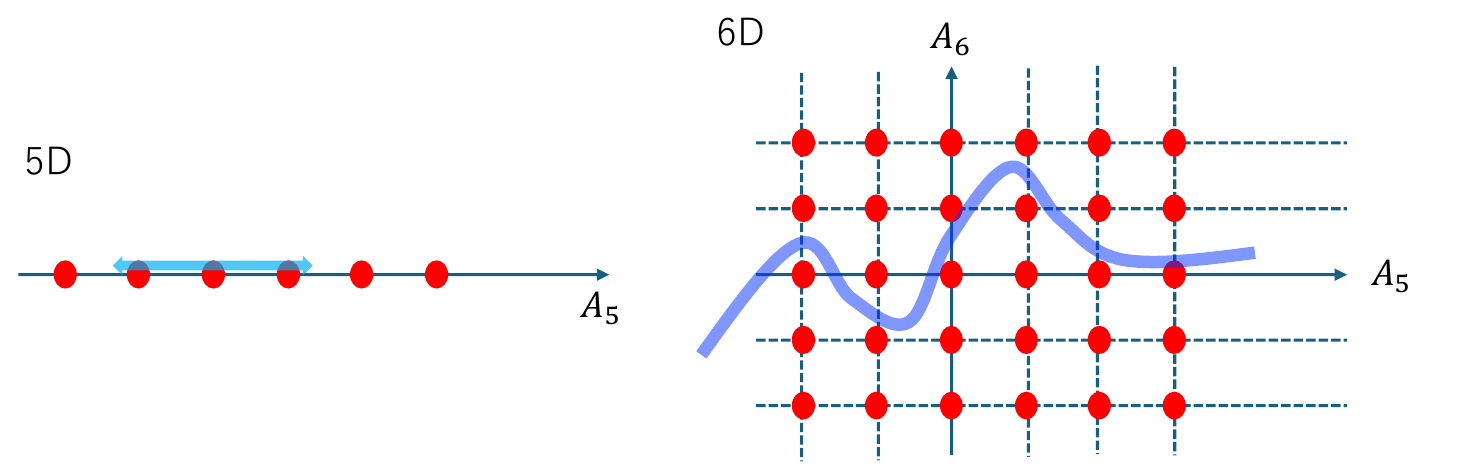}
    \caption{The field excursion of the gauge field(s) in 5D (left) and 6D (right). The red points correspond to the field values at which one of KK masses becomes zero. For 5D cases, the gauge potential necessarily crosses the sweet spots as long as its field excursion is large enough. However, in 6D cases, since each Wilson line phase yields effective KK masses, gauge potential may avoid crossing the sweet-spots as indicated by the blue curve. In such a case, KK particle production is expected to be small if field velocity is small.}
    \label{fig:5D6D}
    \end{figure}
    In 5D scalar QED model, if the variation magnitude $|qg\Delta A_5|$, which can also be understood as the total acceleration/deceleration, is sufficiently large, KK particles are unavoidably produced ~\cite{Yamada:2024aca}. However, in our 6D model, this might not be the case: From \eqref{effKKmass}, the sweet-spots of particle production are 
\begin{align}
    \langle A_6+\ri A_5\rangle=\frac{n_I+\ri n_R}{qg_{4D}R},
\end{align}
at which an effective mass of a KK mode vanishes, and therefore, if $\zeta\sim \langle A_6+\ri A_5\rangle$ move on the complexified field space generally, it is possible that the moduli do not cross these sweet-spots even though the total field excursion is large enough. Such a situation is schematically shown in Fig.~\ref{fig:5D6D}.  In other words, 5D case can be understood as a particular limit $A_6=\frac{m}{qg_{4D}R}$ with $\forall m\in {\mathbb Z}$, which may be realized if the effective potential of $A_6$ stabilizes the modulus at such points. We emphasize that at the sweet-spots, the KK mass becomes precisely zero, and therefore, however heavy they were, the corresponding modes would be non-perturbatively produced. In this sense, if KK Schwinger effect takes place, the 4D effective theory given by the truncation of massive KK modes cannot be justified and one has to leave the KK modes that can be produced via KK Schwinger effect. This does not imply the breakdown of 4D effective theory since only a finite number of KK modes can be produced. Nevertheless, since KK particles namely the modes having momenta along compact directions are produced, there would be e.g. electric currents along compact directions, and therefore, the resultant theory is not just 4D effective theory in the strict sense. Note that if there is no particle production, each sweet-spot in the moduli space gives precisely the same 4D effective theory since only relabeling of KK modes occurs. However, since KK particles are materialized as real particles in the presence of the electric field background, each point in the moduli space no longer gives the same effective theory.
\begin{figure}[htbp]
    \centering
\includegraphics[keepaspectratio, scale=0.8]{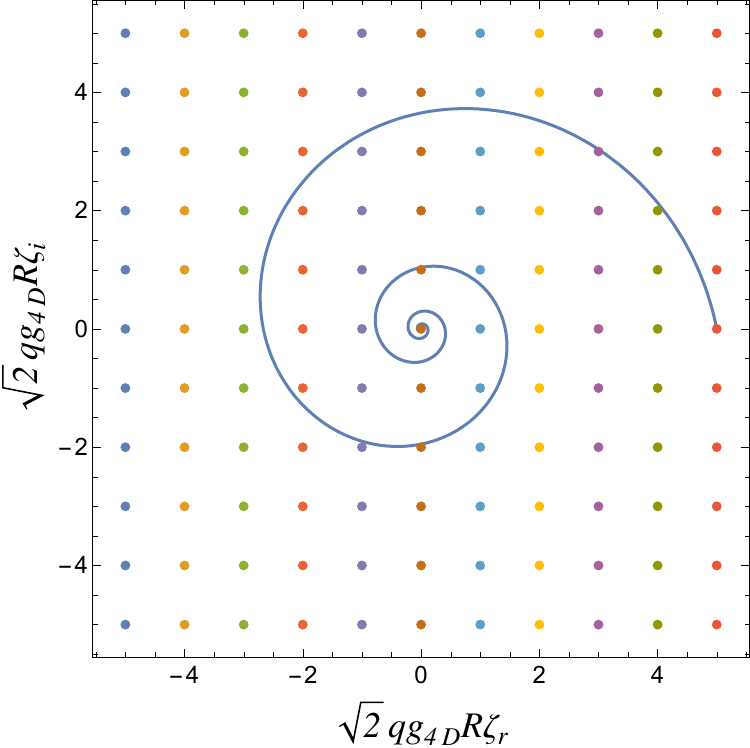}
    \caption{The field trajectory for $A=5$, $m=0.05,\gamma=0.01$ in the unit of $R=1$ with \eqref{KKschwingermodel}. The blobs denote the sweet spots at which corresponding KK particles become massless.}
    \label{fig:zeta-trajectory}
    \end{figure}
    
\begin{figure}[htbp]
    \centering
\includegraphics[keepaspectratio, scale=0.8]{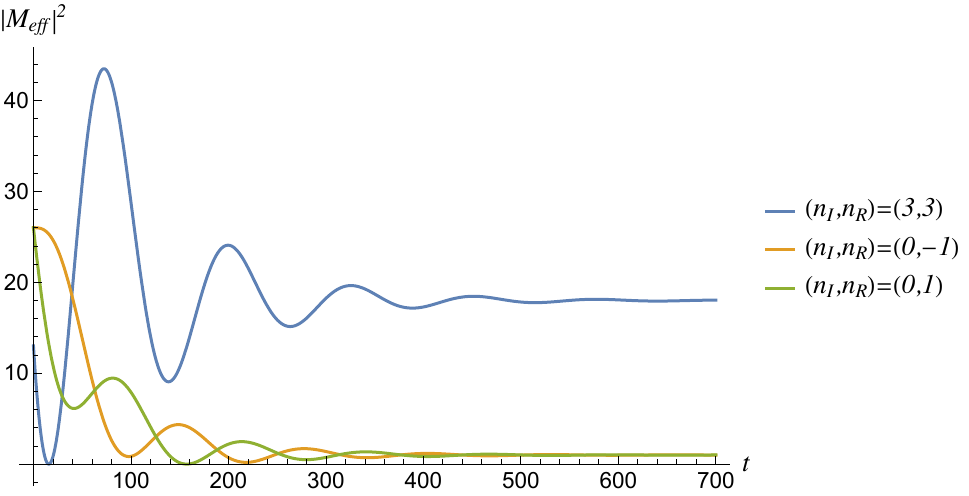}
    \caption{The time dependence of effective masses. We have taken $qg_{\rm 4D}=1$ and the rest same as that in Fig.~\ref{fig:zeta-trajectory}. We have picked up the modes whose effective mass becomes very small compared with its final values. }
    \label{fig:mass}
    \end{figure}

Let us consider an illustrating model where we take the Wilson line modulus to be
\begin{align}
    \zeta(t)= \frac{A}{\sqrt 2qg_{\rm 4D}R}e^{-\gamma t}e^{\ri mt},\label{KKschwingermodel}
\end{align}
where $A$ is an dimensionless parameter and $\gamma,m$ are parameters of mass-dimension 1 describing the decay rate and the ``angular velocity'', respectively. We show the trajectory on the complex $\zeta$ plane in Fig.~\ref{fig:zeta-trajectory}. The colored points correspond to ``sweet-spots'' and the modes that are crossed by the Wilson line modulus would be produced. Notice that we have chosen $m,\gamma$ much smaller than the KK scale $1/R=1$ in the figure, and therefore, the corresponding electric field energy scale is sufficiently smaller than the KK scale. In Fig.~\ref{fig:mass}, we have picked up some KK modes and shown the behavior of time-dependent KK mass for each mode. The time-dependent mass oscillates several times but eventually ends up with the values determined by their KK number. Note that we have named each mode by their KK momentum in the asymptotic future $t\to+\infty$.
    
    \begin{figure}[htbp]
    \centering
\includegraphics[keepaspectratio, scale=0.8]{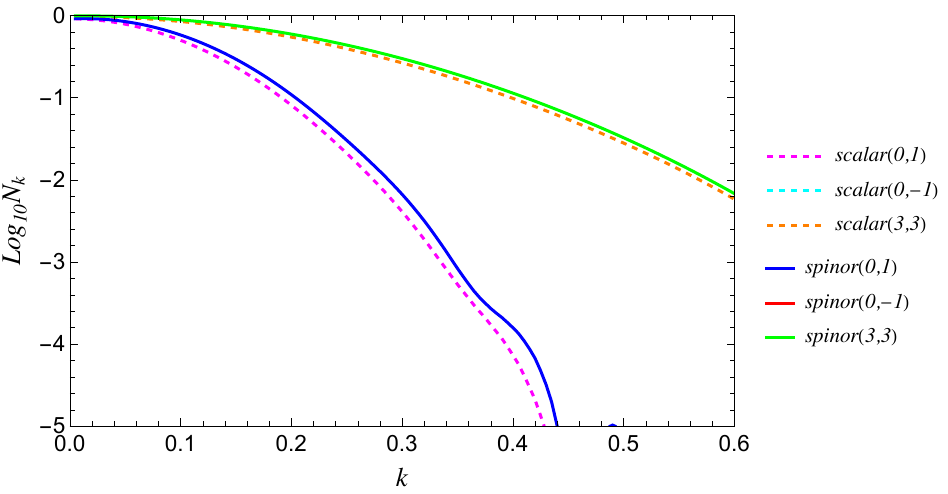}
    \caption{The number density of KK particles. We have taken the parameters in Figs.~\ref{fig:zeta-trajectory},~\ref{fig:mass} and in the unit $R=1$. We have taken the initial time $t_i=0$ and the final time $t_f=700$ at which the particle number density is evaluated. For Dirac spinor, we have chosen $h=+1$. The KK number $(n_I,n_R)$ can be identified as the lattice points in Fig.~\ref{fig:zeta-trajectory}, and the modes that $\zeta(t)$ crosses are indeed produced both for complex scalars and spinors, but the mode $(0,-1)$ which $\zeta$ does not cross is indeed not produced. This figure shows that the creation rate is almost independent of KK numbers as long as $\zeta$ crosses the lattice point. Note that the curves correspond to $(0,-1)$ for scalar and spinor are out of this figure due to the smallness of produced number.}
    \label{fig:kdep}
    \end{figure}

    \begin{figure}[htbp]
    \centering
\includegraphics[keepaspectratio, scale=0.8]{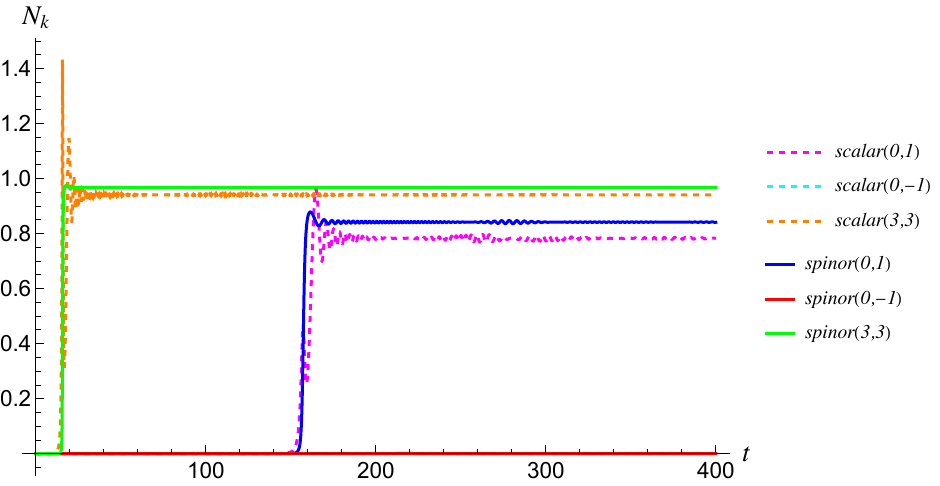}
    \caption{The time-dependence of the number density of KK particles with 3-momentum $k=0.05$ (in the unit $R=1$). We have taken the parameters in Figs.~\ref{fig:zeta-trajectory},~\ref{fig:mass}. As expected, $(3,3)$ modes are first excited and $(0,1)$ modes later, whereas $(0,-1)$ modes which $\zeta$ does not cross are not excited.}
    \label{fig:time}
    \end{figure}

We have solved mode equations of some KK fields both for complex scalars and spinors using Mathematica. In Fig.~\ref{fig:kdep}, we have shown the number density of KK particles with different magnitude of 3-momenta $k$ evaluated $t_f=700$ while the initial time $t_i=0$. We have taken 120 discrete values of $k$ with the unit $k_{\rm unit}=0.005$ and joined the points smoothly in the figure. As expected, the KK modes that $\zeta$ crosses on its complex plane (see Fig.~\ref{fig:zeta-trajectory}) are produced as we expected. Particularly interesting case would be the mode $(n_I,n_R)=(3,3)$, which becomes significantly heavy in the asymptotic future and there is no way to produce such heavy modes within perturbative scattering processes. As also expected, the modes that $\zeta(t)$ does not cross on its field space have no chance to become light, and therefore, are not produced for any small $k$ modes.\footnote{Precisely speaking, such modes can be produced but its amount is negligibly smaller than that of the modes that become massless at some time.} Such a naive expectation can be confirmed by these results. Notice however, if we decrease the parameter $m$ further, only smaller $k$ mode would be produced via this process since the field velocity $\dot\zeta$ is proportional to $m$, which determines the typical momentum of created particle.

The particle-creation time can be estimated as the time at which an effective KK mass vanishes, and is independent of spin of particles as we see in Fig.~\ref{fig:time}. Notice also that the time-dependent particle number density can exceed one for bosons whereas not for fermions by Pauli's exclusion principle. Indeed, if there are multiple particle events, they may interfere with each other and the boson number density may become significantly large as the case of the preheating in the early universe~\cite{Kofman:1997yn} but not for fermions~\cite{Greene:2000ew,Peloso:2000hy}. Such a difference would not be important for the case where a mode experiences particle production at most once,  and this is why we find that the creation patterns of scalars and spinors are almost the same in Figs.~\ref{fig:kdep}, \ref{fig:time}.

What would be the fate of the electric field/Wilson line modulus? As briefly discussed in \cite{Yamada:2024aca}, the back-reaction from the produced particles becomes non-negligible as the particle number density becomes large. Similar situation has been discussed in literature~\cite{Kofman:2004yc,Tanji:2008ku,Kikuchi:2023uqo}. The issue of back-reaction always requires much computational resources and we will not address it here.

Another issue would be the fate of the KK particles produced from vacuum. We note that the electric field configuration depends only on time $t$ and the translational invariance on a torus is unbroken, which implies KK momentum conservation. Therefore, due to the KK momentum conservation, the leading order process to reduce produced KK particles is the coannihilation of the KK particle-anti-particle pairs. Since the momentum distribution of KK (anti-)particle produced via KK Scwhinger effect is fixed only by specifying the time-dependence of the gauge potential $\zeta(t)$, which is quite model-dependent. Nevertheless, the conservation of the KK momentum implies a possibility for KK particles to be a candidate for a dark matter component, which may give constraints on the dynamics of gauge potential $\zeta$. We will leave the study of the relic KK particles for future work.

In summary of this section, we have shown that with pure electric field background, the acceleration/deceleration of KK momenta may lead to non-perturbative particle production, which occur within low-energy effective theory below KK scale. As we have explicitly seen, the trajectory of the Wilson line modulus determines which mode to be produced independently of particle's spin. As we will show, this phenomenon does not take place for the case with magnetic flux background due to the different KK level structure.

\section{Electric background with magnetic fluxes}\label{withmagneticflux}
Let us turn to the case with magnetic flux $N\neq0$. In this case, the KK spectrum of the model differs from that without magnetic fluxes and couplings of the hypermultiplet to the electric background become more involved. 

There are two different methods to treat the time-dependent gauge potential. The first approach that resembles the one in the previous section is as follows. We expand (super)fields by wave functions that do not contain the time-dependent gauge potential and treat the gauge potential as an external potential. It turns out that such a treatment gives an infinite dimensional KK mass matrices which are difficult to diagonalize. Just for completeness, we show such an approach in appendix~\ref{externalpotential}.  Instead we present another approach: We consider Dirac operators containing the time-dependent Wilson line $\zeta(t)$. As we will show below, the eigenvalues of such Dirac operator is independent of the time-dependent Wilson line as is the case of time-independent Wilson line. From such a property, we find that the KK Schwinger effect shown before does not happen in the magnetized model. 

\subsection{Time-dependent Wilson line approach}\label{TDWL}
We here take an approach to derive the 4D effective action in which we expand charged superfields in mode functions that non-perturbatively include the time-dependent gauge potential, which we call a ``time-dependent Wilson line'' approach. 

First, we discuss the mode function without time-dependent Wilson lines. We formally expand the hypermultiplet in KK modes as\footnote{Since $Q$ and $\bar{\tilde Q}$ have the same $U(1)$ charge, their KK expansion with an orthonormal set should coincide with each other, and the complex conjugation implies $\tilde Q$ should be expanded as that shown below. }
\begin{align}
Q(x,z,\bar{z},\theta,\bar\theta)=&\sum_{n,j}Q_{n,j}(x,\theta,\bar\theta)f_{n,j}(z,\bar{z}),\\
\tilde{Q}(x,z,\bar{z},\theta,\bar\theta)=&\sum_{n,j}\tilde{Q}_{n,j}(x,\theta,\bar\theta)\bar{f}_{n,j}(z,\bar{z}),
\end{align}
where the label $n$ denotes the KK quantum number and $j=1,\cdots, |N|$ denotes the $|N|$-degenerate modes, which will be discussed below. The KK mode functions $f_{n,j}(z,\bar{z})$ and $\bar{f}_{n,j}(z,\bar{z})$ are defined as follows: One can assume $N>0$ without loss of generality. Consider the Dirac operators\footnote{By performing the $\theta^2$ integration of the superpotential, one finds the differential operators indeed act on the fermionic components of $Q$ and $\bar{\tilde Q}$, respectively.} on $\mathbb{T}^2$
\begin{align}
    \mathcal{D}\equiv \partial-\frac{\pi N\bar{z}}{2},\quad
    \bar{\mathcal{D}}\equiv -\left(\bar\partial+\frac{\pi N z}{2}\right),
\end{align}
where the former appears in the action as the operator acting on $Q$ whereas the latter as the operator acting on $\bar{\tilde Q}$ (after integration by part). One finds the following set of creation and annihilation operators defined from the above differential operators
\begin{align}
    \hat{a}\equiv \frac{1}{\sqrt{\pi N}}\bar{\cal D},\quad \hat{a}^\dagger\equiv \frac{1}{\sqrt{\pi N}}\cal D
\end{align}
which satisfies $[\hat{a},\hat{a}^\dagger]=1$. It is known that there is a normalizable zero mode function $f_{0,j}(z,\bar{z})$ satisfying $\hat{a}f_{0,j}(z,\bar{z})=0$ and given by
\begin{align}
    f_{0,j}(z,\bar{z})=\mathcal{N}e^{\ri Nz\frac{{\rm Im}z}{{\rm Im }\tau}}\vartheta\left[\begin{array}{c}\frac{j}{N}\\ 0\end{array}\right]\left(N z,\ri N\right),\label{magzeromode}
\end{align}
where $\vartheta\left[\begin{array}{c}a\\ b\end{array}\right](\nu,\tau)\equiv \sum_{l\in\mathbb{Z}}\exp\left(\pi(a+l)^2\tau+2\pi \ri(a+l)(\nu+b)\right)$ is the Jacobi theta function and $\mathcal{N}$ is a normalization factor such that $\int d^2z\sqrt{g_2}|f_{0,j}(z,\bar{z})|^2=1$ where $\sqrt{g_2}$ is the volume element on torus. Note that there are $|N|$-degenerate zero modes labeled by $j=1,\cdots,|N|$. The higher KK mode functions can be constructed from the zero mode function by multiplying $\hat{a}^\dagger$ as
\begin{align}
   f_{n,j}(z,\bar{z})=\frac{1}{\sqrt{n!}} (\hat{a}^\dagger)^n f_{0,j}(z,\bar{z}),
\end{align}
which satisfy the orthonormal condition
\begin{align}
    \int d^2z \sqrt{g_2}\bar{f}_{m,j_1}f_{n,j_2}=\delta_{j_1j_2}\delta_{mn}.
\end{align}
Thus, we are able to derive 4D effective action with the aid of the above properties.\footnote{The explicit forms of wave functions can be found e.g. in~\cite{Hamada:2012wj}.}

Let us introduce time-dependent Wilson line $\zeta(t)$. An important observation is that we are able to find eigenfunctions of a Dirac operator
\begin{align}
    \mathcal{D}_\zeta=\partial-\frac{\pi N\bar z}{2}-\frac{g q\zeta(t)}{\sqrt2}\equiv \partial-\frac{\pi N(\bar z+\bar{\zeta}_W(t))}{2}
\end{align}
where we have introduced a new notation ${\bar\zeta}_W(t)\equiv \frac{\sqrt2 gq\zeta(t)}{\pi N}$. Now, one can regard $\bar\zeta_W(t)$ as a (time-dependent) shift of the coordinate, and notice that $\mathcal{D}_\zeta$ satisfies the same algebra as that without $\zeta$. The eigenfunction of the Dirac operator is known when $\zeta$ is a constant and is generalized to the case with $\bar\zeta_W(t)$ as
\begin{align}
    f_{0,j}^\zeta(z,\bar{z};t)=&\mathcal{N}e^{\ri N(z+\zeta_W(t)){\rm Im}(z+\zeta_W(t))}\vartheta\left[\begin{array}{c}\frac{j}{N}\\ 0\end{array}\right]\left(N (z+\zeta_W(t)),\ri N\right),\\
    f_{n,j}^\zeta(z,\bar{z};t)=&\frac{1}{\sqrt{n!}}(\hat{a}_\zeta^\dagger)^nf^\zeta_{0,j}(z,\bar{z};t),
\end{align}
where $\hat{a}_\zeta^\dagger\equiv \frac{1}{\sqrt{\pi N}}{\cal D}_\zeta$. These are straightforward generalization of the case without $\zeta$ and the eigenvalues of Dirac and Laplace operator are exactly the same as that case. Therefore, the ``KK mass'' originating from the Laplace operator does not depend on the time-dependent Wilson line. 

Although the eigenvalues of Dirac and Laplace operator on ${\mathbb T}^2$ are independent of time-dependent Wilson line, electric field along compact directions lead to additional effective mass terms that differ from the KK mass terms given by the eigenvalues of Dirac and Laplace operators: Integration over compact spaces can be done straightforwardly since the time-dependent Wilson line $\zeta_W(t)$ is simply the shift of the coordinates. However, we notice that time-derivatives in the components of superfields now act on both 4D fields and the wave function in the KK expansion of each field. The latter yields additional time-dependent mass (mixing) terms other than the usual KK mass terms obtained by diagonalizing the extra-dimensional (covariant) kinetic terms. To extract such additional mass terms in the present superspace, we decompose the components of $Q$ and $\tilde{Q}$ as
\begin{align}
Q(x,\theta,\bar\theta,z,\bar{z})=&\sum_{n,j}Q_{n,j}(x,\theta,\bar\theta)f_{n,j}^\zeta(z,\bar{z};t)
    +\delta Q(x,\theta,\bar\theta,z,\bar{z}),\\
    \tilde{Q}(x,\theta,\bar\theta,z,\bar{z})=&\sum_{n,j}\tilde{Q}_{n,j}(x,\theta,\bar\theta)\bar{f}^\zeta_{n,j}(z,\bar{z};t)
    +\delta \tilde{Q}(x,\theta,\bar\theta,z,\bar{z}).
\end{align}
Here we have defined
\begin{align}
    \delta Q(x,\theta,\bar\theta,z,\bar{z})=&\sum_{n,j}\Biggl[\ri\theta\sigma^0\bar\theta Q_{n,j}(x)\dot{f}^\zeta_{n,j}-\frac{\ri}{\sqrt2}\theta^2\dot{f}^\zeta_{n,j}\psi_{n,j}(x)\sigma^0\bar\theta+\frac14\theta^2\bar\theta^2\partial_0(Q_{n,j}(x)\dot{f}^\zeta_{n,j})\Biggr]\nonumber\\
    =&\sum_{n,j}\delta Q_{n,j}(x,\theta,\bar\theta) \dot{f}^\zeta_{n,j}(z,\bar{z};t)+\mathcal{O}(\ddot{f}^\zeta_{n,j}),
\end{align}
where we have applied the adiabatic expansion with respect to $\zeta$ on the second line and
\begin{align}
    \delta Q_{n,j}(x,\theta,\bar\theta)\equiv\ri\theta\sigma^0\bar\theta Q_{n,j}(x)-\frac{\ri}{\sqrt2}\theta^2\psi_{n,j}(x)\sigma^0\bar\theta+\frac14\theta^2\bar\theta^2\dot{Q}_{n,j}(x).
\end{align} 
In essence, $\delta Q$ refers to the terms that time-derivatives act on wave function $f^\zeta_{n,j}$. One can similarly find $\delta Q^c$, which appears due to the time-dependence of mode functions. Notice that 
\begin{align}
    \dot{f}_{n,j}^\zeta(z,\bar{z};t)=(\dot\zeta_W\partial+\dot{\bar\zeta}_W\bar\partial)f^\zeta_{n,j}(z,\bar{z};t)
\end{align}
since $\zeta_W$ appears as the shift of the $z$-coordinate. In the weak electric field limit $\dot\zeta\ll \frac{1}{\sqrt{\cal A}}$, the leading order correction to the quadratic action is given by
\begin{align}
  \delta\mathcal{S}_2^\zeta &= \int d^6X\sqrt{-G}\int d^4\theta\left[ \bar{Q}\delta Q+\delta{\bar Q}Q\right]+{\mathcal{O}(\ddot\zeta_W,\dot{\zeta}_W^2)}\nonumber\\
    &\approx\int d^6X\sqrt{-G}\int d^4\theta\sum_{n,m}\left[\bar{Q}_{n,j}\delta Q_{m,j}\bar{f}^\zeta_{n,j}(\dot\zeta_W\partial+\dot{\bar\zeta}_W\bar\partial)f^\zeta_{m,j}+\delta\bar{Q}_{n,j} Q_{m,j}(\dot\zeta_W\partial+\dot{\bar\zeta}_W\bar\partial)\bar{f}^\zeta_{n,j}f^\zeta_{m,j}\right]\nonumber\\
    &=\frac12\int d^6X\sqrt{-G}\int d^4\theta\sum_{n,m}\left[\bar{Q}_{n,j}\delta Q_{m,j}\{\bar{f}^\zeta_{n,j}(\dot\zeta_W\mathcal{D}_\zeta-\dot{\bar\zeta}_W\bar{\cal D}_\zeta)f^\zeta_{m,j}-(\dot\zeta_W\mathcal{D}_\zeta-\dot{\bar\zeta}_W\bar{\cal D}_\zeta)\bar{f}^\zeta_{n,j}f^\zeta_{m,j}+{\rm h.c.}\}\right]\nonumber\\
    &=\frac{\sqrt{\pi N}}{2}\int d^6X\sqrt{-G}\int d^4\theta\sum_{n,m}\left[\bar{Q}_{n,j}\delta Q_{m,j}\{\bar{f}^\zeta_{n,j}(\dot\zeta_W\hat{a}_\zeta^\dagger-\dot{\bar\zeta}_W \hat{a}_\zeta)f^\zeta_{m,j}+(\dot\zeta_W\hat{a}_\zeta-\dot{\bar\zeta}_W\hat{a}_\zeta^\dagger)\bar{f}^\zeta_{n,j}f^\zeta_{m,j}\}+{\rm h.c.}\right]\nonumber\\
    &=\sum_{n}\frac{2\sqrt2 M_{n}g_{\rm 4D}q}{M_0^2}\int d^4x\int d^4\theta\left[\dot{\bar{\zeta}}(t)\bar{Q}_{n+1,j}\delta Q_{n,j}-\dot{\zeta}(t)\bar{Q}_{n,j}\delta Q_{n+1,j}+{\rm h.c.}\right],
\end{align}
where $M_n\equiv \sqrt{\frac{4\pi N(n+1)}{\mathcal{A}}}$, we have performed integration by part in the second equality to make covariant derivatives and note that $\hat{a}_\zeta=-{\cal D}_{\zeta}$ and $\hat{a}^\dagger_\zeta=-\bar{\cal D}_\zeta$ when it acts on $\bar{f}^\zeta_{n,j}$. Thus, we have found the leading order corrections to the quadratic action due to the electric field. 

Unlike the case without magnetic field, KK Schwinger effect is not expected in this case: The above level mixing terms that originate from the time-dependence of the wave function are proportional to the electric field not the time-dependent Wilson line itself. Recall that the KK masses originating from Dirac and Laplace operators are independent of electric field. Therefore, the effective KK mass matrices consist of time-independent KK mass terms as diagonal elements and the level mixing terms as off diagonal elements. Then, the level crossing may take place only when the electric field is at least compatible with the lowest KK scale $M_0= \sqrt{4\pi N/\cal A}$, which is inconsistent with the energy condition for the 4D effective field theory $E<M_0$. Thus, we conclude that KK particle production such as the KK Schwinger effect does not take place for magnetized models within 4D effective theory unlike the case without magnetic fluxes. Such a difference originates from the KK level structure of the magnetized model. We give a simple and intuitive explanation for the magnetized model in the next subsection within a quantum mechanical system. 

Despite the absence of KK Schwinger effect, the electric field on torus may affect low energy 4D physics when one considers interactions of bulk zero modes that non-perturbatively couple to the time-dependent Wilson lines. We will discuss its physical effect in Sec.~\ref{TDWL10D} by generalizing our results to 10D super Yang-Mills model, which yields a semi-realistic matter spectrum. 

\subsection{Quantum Hall effect and electric field}
In order to have a better intuition about the physics of electric field on a magnetized torus, we discuss the role of electric field in the context of quantum Hall effect, which captures the essence of the electric field in magnetized models. We briefly review it on the basis of~\cite{Tong:2016kpv}. Consider an electrically charged particle in 2D space with Hamiltonian
\begin{align}
    \hat{H}=\frac{1}{2m}\left(\hat{\bm p}+e{\bm A}(\hat{\bm x})\right)^2.
\end{align}
Magnetic field along $z$-direction (orthogonal to the 2D surface we consider) is realized by taking
\begin{align}
    {\bm A}(\hat{\bm x})=B\left(\begin{array}{c}0\\ \hat{x}\end{array}\right),
\end{align}
where $B$ is the magnetic field strength and the Hamiltonian becomes
\begin{align}
    \hat{H}=\frac{1}{2m}\left(\hat{p}_x^2+(\hat{p}_y+eB\hat{x})^2\right)
\end{align}
The position space wave function is expanded as
\begin{align}
    \Psi(x,y)=\sum_{k}e^{\ri k y}f_k(x),
\end{align}
where we have introduced a formal sum of $k$ but it should be replaced by integral for continuous momentum case. The effective Hamiltonian for $p_y=k$ mode is 
\begin{align}
    \hat{H}_k=\frac{1}{2m}\left(\hat{p}_x^2+(k+eB\hat{x})^2\right)=\frac{1}{2m}\hat{p}_x^2+\frac{m\omega_B^2}{2}\left(\hat{x}+\frac{k}{eB}\right)^2,
\end{align}
where $\omega_B=eB/m$ is the cyclotron frequency. Thus, the system can be regarded as harmonic oscillator for a fixed $k$, and energy level is quantized as
\begin{align}
    E_{n,k}=\omega_B\left(n+\frac12\right),
\end{align}
which corresponds to the KK number in the magnetized torus models.

Now, we introduce constant electric field along $x$-direction by adding electric potential $\phi=-Ex$
\begin{align}
    \hat{H}=\frac{1}{2m}\left(\hat{p}_x^2+(\hat{p}_y+eB\hat{x})^2\right)+eE\hat{x}.
\end{align}
Again, we consider a wave function having $p_y=k$, and its effective Hamiltonian becomes
\begin{align}
    \hat{H}_k&=\frac{1}{2m}\hat{p}_x^2+\frac{m\omega_B^2}{2}\left(\hat{x}+\frac{k}{eB}\right)^2+eE\hat{x}\nonumber\\
    &=\frac{1}{2m}\hat{p}_x^2+\frac{m\omega_B^2}{2}\left(\hat{x}+\frac{k}{eB}+\frac{mE}{eB^2}\right)^2-\frac{kE}{B}-\frac{E^2m}{2B^2}.
\end{align}
Therefore, the energy level is given by
\begin{align}
    \tilde{E}_{n,k}=\omega_B\left(n+\frac12\right)-\frac{kE}{B}-\frac{E^2m}{2B^2},
\end{align}
which indicates that the Landau level structure is intact under the constant electric field since the shift of energy is uniform for any $n\in\mathbb{N}\cup\{0\}$. 

Despite differences of some technical points such as gauge conditions and spatial periodic boundary conditions, the above quantum mechanical system explains KK level structure on magnetized torus. In our case of the time-dependent electric field realized by time-dependent Wilson lines $\zeta_W(t)$, the gauge field configuration differs from what we have shown here. However, the above result helps understanding of the physics on magnetized torus: The electric field/time-dependent Wilson lines shift the KK tower uniformly. This observation is consistent with the fact that the time-dependent Wilson lines do not change the eigenvalues of Dirac operators. Furthermore, as expected from the above harmonic oscillator description, the peak of the Gaussian-like wave function would move due to the electric gauge potential and indeed, the time-dependent Wilson lines lead to the shift of the peak of the zero mode wave functions, which causes the time-dependent 4D effective couplings as discussed in the next subsection.

\subsection{Electric field in magnetized 10D super Yang-Mills theory and flavor structure}\label{TDWL10D}
We have found that the electric field does not cause KK Schwinger effect within magnetized models because of the KK level structure, and therefore as long as we are concerned with the physics below the KK scale, no KK particles are produced and the truncation of heavy fields are consistently achieved. Then, one may wonder whether the electric field would affect the low energy physics containing only zero modes. Indeed, the couplings in 4D effective theory depend on zero mode wave functions affected by time-dependent Wilson lines. One of the most important effects is on the Yukawa couplings within magnetized models: In the magnetized models, there appear some number of zero modes, which can be identified with ``generation'' in the 4D effective theory. The flavor structure is then determined by how the zero mode wave functions are localized on tori as the 4D effective Yukawa couplings are determined by the overlap integrals of the wave functions of matter fields. Since the time-dependent Wilson lines (non-uniformly) translate the wave functions and change their overlap integrals, we expect that the flavor structure crucially depends on the time-dependent Wilson lines/electric fields.

In order to discuss how the time-dependent Wilson lines affect the low energy physics within a semi-realistic setup, we need to generalize our discussion to super Yang-Mills theory with space dimensions more than five, namely 8D or 10D super Yang-Mills theory. We here consider 10D case and the semi-realistic model can be derived as follows: The U(1) gauge theory is replaced by a larger gauge group such as $U(8)$ that is broken to $SU(3)\times SU(2)$ with multiple $U(1)$ subgroups via introduction of Abelian magnetic fluxes or Wilson lines. Such a semi-realistic model was proposed  in~\cite{Abe:2012ya,Abe:2012fj}. Now, the ``matter hypermultiplets'' in 6D models are replaced by off-diagonal components of adjoint representations in $U(N)$, particularly the gauge potential along compact directions and their superpartners in 4D effective theory, which become the matter fields of the standard model sector.\footnote{In this sense, some off-diagonal components of $U(N)$ become matter whereas the diagonal components are Wilson line moduli. Note that there also appear exotic matter in the off-diagonal part (see appendix~\ref{10Dreview} for more details). } As we will see below, the generalization of such models to include electric fields can be achieved simply by replacing the constant Wilson lines with time-dependent ones.

We here discuss the behavior of Yukawa couplings that contain time-dependent Wilson lines. For completeness, we have reviewed the derivation of semi-realistic model and the Yukawa couplings in appendix~\ref{10Dreview} and the definition of each quantity is shown there as well. In the model of \cite{Abe:2012fj}, there appear six generation of both up and down type Higgs fields which is a consequence of three generation structure for quarks and leptons, and the Ansatz of the VEV of those Higgs fields are taken as
\begin{align}
    \langle H^K_u\rangle=&(0.0,0.0,2.7,1.3,0.0,0.0)v\sin\beta\times N_{H_u},\\
    \langle H^K_d\rangle=&(0.0,0.1,5.8,5.8,0.0,0.0)v\cos\beta\times N_{H_d},\\
    \tan\beta=&25,
\end{align}
where the normalization factors are $N_{H_u}=(2.7^2+1.3^2)^{-\frac12}$, $N_{H_d}=(0.1^2+5.8^2+5.8^2)^{-\frac12}$. A set of the Wilson lines and the VEV of complex structure modulus on the first torus are chosen as
\begin{align}
&(\zeta_Q,\zeta_U,\zeta_D,\zeta_L,\zeta_N,\zeta_E)=(1.0\ri,1.9\ri,1.4\ri,0.7\ri,2.2\ri,1.7\ri),\\
&\tau_1=4.1\ri,
\end{align}
which are responsible for the flavor structure such as mass hierarchy. (See \cite{Abe:2012fj} for details and the resultant flavor structure with these Ansatz.)\footnote{We note that the above Ansatz yields Yukawa couplings at the scale $10^{16}$ GeV and after including renormalization group running, one finds semi-realistic values of fermion masses and mixing angles. See~\cite{Abe:2012fj} for more details.} Other parameters such as 10D gauge coupling $g$ and area of each torus $\mathcal{A}_{1,2,3}$ fix the overall factor of Yukawa couplings, and does not change the flavor structure such as mass ratio. In \cite{Abe:2012fj}, those parameters are given in terms of the VEV of moduli superfields rather than the area or gauge coupling. We here choose parameters different from that in \cite{Abe:2012fj} as
\begin{align}
    &\mathcal{A}_i=1, \quad \forall i=1,2,3,\\
    &g=1.68,
\end{align}
which are chosen just for an illustrating purpose.

\begin{figure}[htbp]
    \centering
\includegraphics[keepaspectratio, scale=1]{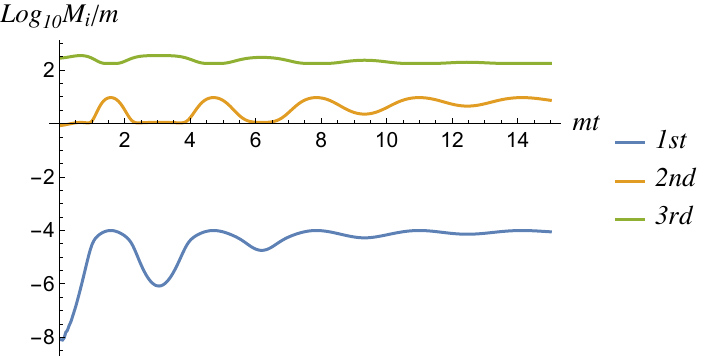}
    \caption{The time-dependence of the up-type quark mass eigenvalues with the time-dependent Wilson line. The parameters are chosen as $m=1{\rm GeV}$, $\gamma/m=0.1$.}
    \label{fig:Yukawa}
    \end{figure}

Let us discuss how the flavor structure may change by the time-dependent Wilson lines on tori. We note that the flavor structure within our example originates from the wave functions on the first torus, and therefore, we consider the case where the Wilson lines on the first torus are turned on. Notice  that even if the Wilson lines on the second or third torus are turned on, the flavor structure would not be affected, which would be clear from the derivation of Yukawa couplings shown in appendix~\ref{10Dreview}. Nevertheless, the time-dependent Wilson line on the 2nd and 3rd torus may cause KK Schwinger effect for the fields that feel zero effective flux as is the case discussed in Sec.~\ref{pureelectric}. As an simple illustration, we replace the Wilson line shown in the above by
\begin{align}
    \zeta_Q\to\zeta_Q(t)=1.9\ri+5\ri e^{-\gamma t} e^{-\ri mt}\label{zetamodel}
\end{align}
which results in the time-dependent flavor structure for quarks. One can substitute this background as well as other parameters into \eqref{Yukawa} and find a formula for the fermion mass matrices. We show the behavior of the eigenvalues of up type quark mass matrix in Fig.~\ref{fig:Yukawa}, which shows that the time-dependence of the Wilson line affects the mass of the lightest generation (1st generation) most significantly. Here we have assumed that the dimensionless quantity $\zeta_Q$ to be $\mathcal{O}(1)$, but what would be the necessary ``electric energy''? Recalling that 
\begin{align}
    \langle A^{aa}_{z^i}\rangle=-\frac{\pi M_a^{(i)}}{2{\rm Im}\tau_i}(\bar{z}^i+\bar{\zeta}^i_a)\to F_{0z}=\langle \dot{A}^{aa}_{z^i}\rangle=-\frac{\pi M_a^{(i)}}{2{\rm Im}\tau_i}\dot{\bar{\zeta}}^i_a,
\end{align}
the electric field energy density (on the $j$ th torus) in 4D effective theory is
\begin{align}
  \rho_{E_j^a}=& \prod_{i=1}^3 \int d^2z^i\sqrt{g_i} \frac{1}{2g^2}F^{aa}_{0z^j}h^{z^j\bar{z}^j}F^{aa}_{0\bar{z}^j}=\frac{{\cal A}_1{\cal A}_2{\cal A}_3}{g^2(2\pi R_j)^2}\frac{\pi^2 (M_a^{(j)})^2}{4({\rm Im}\tau_j)^2}(\dot{\bar{\zeta}}^j_a)^2\nonumber\\
  =&\frac{\pi^2(M_a^{(j)})^2}{4g_{\rm 4D}^2{\cal A}_j{\rm Im}\tau_j}(\dot{\bar{\zeta}}^j_a)^2
\end{align}
where we have introduced $g^2_{\rm 4D}=g^2/(\prod_i{\cal A}_i)$ is the gauge coupling in 4D effective theory.\footnote{This should be the value at the GUT scale and the renormalization group running has to be taken account of to get the gauge couplings at low energy.} Thus, for our model~\eqref{zetamodel}, as long as the parameters satisfy $m,\gamma<M_{\rm KK}^{(j)}=1/\sqrt{{\cal A}_j}$, the energy scale is below the one for the effective theory to be valid even if the amplitude of $\zeta$ is ${\cal O}(1)$, where we have introduced $M_{\rm KK}^{(j)}$ as the KK scale characterizing $j$-th torus. Namely, the oscillation of Yukawa couplings (or equivalently fermion masses) may take place within 4D effective theory where the energy scale is below the KK scales. 

Let us briefly discuss the implication of the above observation to physics in the early universe within the 10D super Yang-Mills model: In general, light scalars fluctuate over their field space during inflation. The Hubble friction traps such light fields until the Hubble expansion rate becomes as small as their masses. Therefore, Wilson line scalars if light can also be trapped at some point in field space. As we will give comments below, the Wilson line scalars have masses below the electroweak scale. Therefore, they would start moving after electroweak phase transition. Since fermions would already be massive due to the electroweak symmetry breaking, the motion of Wilson line scalars change fermion masses as we have shown above. Therefore, until all the Wilson line scalars stop, the standard model fermion masses would vary in time. It would be worth studying whether such time varying masses can be consistent with the cosmological observations so far, or yield any interesting physical phenomena. Note also that the time variation of the fermion masses would affect electroweak sphaleron process and therefore may affect baryogenesis processes. We leave these issues for future study. Notice also that the Wilson line moduli behave as (pseudo) Nambu-Goldstone modes and would be very light. Therefore, the field excursion mentioned above would be a rather general consequence, which may lead to serious problems within the magnetized super Yang-Mills models.\footnote{One of the ways to avoid above problems would be to consider orbifolds so that the Wilson line moduli are all projected out, but it is in general difficult to remove all. (See the comments in appendix~\ref{10Dreview}.)}

As a side remark, we find that fermion masses depend on the values of Wilson lines, which implies that the Wilson line modes known as open string moduli may acquire potential due to the one-loop correction from the standard model fermions. Since the fermion mass is proportional to Higgs VEV, the potential would be very small in general. For the case of U(1) gauge theory, magnetic flux leads to a Nambu-Goldstone mode~\cite{Buchmuller:2016gib,Honda:2019ema,Hirose:2024vvx} that is expected never to acquire any mass.\footnote{Explicit computation of loop corrections is performed in~\cite{Honda:2019ema,Hirose:2024vvx}.} The generalization to the non-Abelian gauge theory with Abelian magnetic fluxes has been studied in \cite{Hirose:2019ywp} where $SU(2)$ gauge theory with magnetic flux corresponding to one generator has been introduced. In such a case, it has been shown that the Wilson line mode corresponding to the broken generator becomes a Nambu-Goldstone mode at least at one-loop order. The observation here suggests that some Wilson lines may obtain their mass e.g. from Coleman-Weinberg potential from bosonic and fermionic loops for non-Abelian models with Abelian magnetic fluxes.\footnote{Essentially, if values of Wilson lines change fermion masses or Yukawa couplings, such modes should acquire a mass from quantum corrections. In other words, the Wilson line modes that appear in Yukawa couplings cannot be identified as Nambu-Goldstone modes in this sense. If the Wilson line modes changing Yukawa couplings were true Nambu-Goldstone modes, their field values should not change physical couplings. Therefore, we do not expect that all the Wilson line modes in diagonal parts of non-Abelian gauge fields to be perturbatively massless fields like the Wilson lines of U(1) gauge symmetries. } It would be interesting to further investigate the number of true Nambu-Goldstone modes within non-Abelian gauge theory with magnetic fluxes, which would be more relevant to semi-realistic model buildings.\footnote{We expect that there are similar couplings including sfermions, which may also yield potential to the moduli.} It would be worth investigating the effective mass of open string moduli given by the one-loop corrections as well as whether such light moduli are consistent with current experimental and observational data, which are also left for future investigation.

\section{Conclusion}\label{conclusion}
In this work, we have investigated the role of electric fields within toroidal compactification model on ${\mathbb T}^2$ or its higher dimensional extension. Such electric fields would be turned on when the gauge potential along compact spaces, which often become light moduli fields in 4D effective theory, have time-dependence. In other words, we have discussed the phenomena associated with such moduli fields, which would be more relevant to physics in the early universe. We have focused particularly on the differences between the models with or without magnetic flux, which changes the structure of KK level significantly. Indeed, we have shown that the low energy phenomena associated with electric fields are quite different for models with/without magnetic background.

Without magnetic fluxes, the electric acceleration/deceleration cause the shift of KK level by which KK modes at given time become lighter and are eventually produced by the KK Schwinger effect. The significance of the KK Schwinger effect is that the electric field energy can be small enough to create KK modes since it occurs at the time when the time-dependent KK mass crosses zero because of deceleration of KK momentum. Unlike the 5D case, in 6D models, a general motion of moduli fields on their field space may not cause significant KK particle production since the sweet-spots of KK Scwhinger effect spread over the 2D momentum lattice. Nevertheless, we have shown that KK Schwinger effect produces KK particles and then truncation of KK modes at an initial time is no longer justified due to the production of the KK modes, which become light at a later time.

In the models with the magnetic flux background, it turns out that time-dependent gauge potential leading to electric field does not cause KK Scwhinger effect unlike the case without magnetic fluxes. As we have explained by a quantum mechanical system, the Landau level structure caused by magnetic flux is not mixed by electric field.\footnote{More precisely speaking, the eigenvalues of Dirac operator is independent of the time-dependent Wilson lines.} Therefore, KK particles are unlikely to be produced as long as electric field energy is below KK scale. Nevertheless, we have shown that the gauge field along compact spaces affects low energy physics as a time-dependent Wilson line, which changes the wave function on torus. As the effective coupling constants in 4D effective theory are given by the overlap integral of zero mode wave functions, we have seen that the time-dependent Wilson line results in the time-dependent Yukawa couplings. 

Both cases require more detailed investigations of phenomenological aspects. In particular, it would be interesting to study cosmological aspects of our models. We emphasize that the cases with and without magnetic fluxes could coexist in realistic models. Indeed, within the 10D super Yang-Mills model~\cite{Abe:2012fj,Abe:2012ya}, there appear tori on which effective magnetic flux disappears while all matter fields couple to nonvanishing magnetic fluxes on the first torus. In such cases, phenomena caused by electric fields with/without magnetic backgrounds may occur simultaneously. We leave further study of those phenomena for future work.

Other related issues are in order: In \cite{Condeescu:2017nmo}, the effect of constant electric field within toroidal compactification is examined within string theory by using world sheet calculations. Such an investigation would be complementary to our work where a fully field theoretical approach has been carried out. String theoretical calculation requires corresponding conformal field theory, which is often difficult to be formulated on non-trivial backgrounds whereas the field theory approach cannot reach to the stringy corrections. Nevertheless, it would be possible to investigate further from D-brane picture which may be described within field theory. Another issue we leave is the non-perturbative effects such as D-brane instantons~\cite{Blumenhagen:2006xt,Ibanez:2006da}. In Sec.~\ref{TDWL10D}, we have assumed electroweak symmetry breaking by giving expectation value to some Higgs fields, which would be realized both by supersymmetry breaking soft mass as well as $\mu$-terms in superpotential, which are forbidden within this framework. Nevertheless, $\mu$-terms can be non-perturbatively produced e.g. from D-brane instantons, and the effective size of the $\mu$-terms are determined by the overlap integral of zero mode wave functions of Higgs fields as well as the instantons. Therefore, when time-dependent Wilson lines are turned on, we expect the $\mu$-terms would also be affected, which would results in the modulation of Higgs VEV. Similar effects arise also for Majorana mass terms of right-handed neutrino multiplets within the magnetized models. We leave these issues for future investigation as well.

\section*{Acknowledgement}
YY would like to thank Yuya Ominato for explanation of quantum Hall effect with electric field. We also thank Junichiro Kawamura for discussion about the annihilation processes of KK particles.
\appendix
\section{Notation}\label{notation}
Throughout this work, we use the convention in \cite{Wess:1992cp} where the 4D Pauli matrices are given by
\begin{align}
    \sigma^0=\left(\begin{array}{cc}
    -1&0\\
    0&-1\end{array}\right),\quad \sigma^1=\left(\begin{array}{cc}
    0&1\\
    1&0\end{array}\right),\quad \sigma^2=\left(\begin{array}{cc}
    0&-\ri\\
    \ri&0\end{array}\right),\quad \sigma^3=\left(\begin{array}{cc}
    1&0\\
    0&-1\end{array}\right),
\end{align}
$\bar{\sigma}^0=\sigma^0$ and $\sigma^i=-\bar{\sigma}^i$ ($i=1,2,3$). These definitions differ from that in \cite{Dreiner:2008tw} where various computational techniques for two component spinors are summarized. In adopting Dreiner-Haber-Martin (DHM) results~\cite{Dreiner:2008tw} for the Wess-Bagger(WB) one, one has to use the following relation:
\begin{align}
   & (\sigma^{\rm WB}_\mu)_{\alpha\dot\beta}=(\sigma^{\rm DHM}_0)_{\alpha\dot\gamma}(\bar{\sigma}^{\rm DHM}_\mu)^{\dot\gamma\delta}(\sigma^{\rm DHM}_0)_{\delta\dot\beta},\\
   & (\bar{\sigma}^{\rm WB}_\mu)^{\dot\alpha\beta}=(\bar{\sigma}^{\rm DHM}_0)^{\dot\alpha\gamma}(\sigma^{\rm DHM}_\mu)_{\gamma\dot\delta}(\bar{\sigma}^{\rm DHM}_0)^{\dot\delta\beta}
\end{align}
or equivalently 
\begin{align}
    (\sigma_\mu^{\rm DHM})_{\alpha\dot\beta}=(\sigma_0^{\rm WB})_{\alpha\dot\gamma}(\bar{\sigma}_\mu^{\rm WB})^{\dot\gamma\delta}(\sigma_0^{\rm WB})_{\delta\dot\beta}
\end{align}
with the metric convention $(+---)$ in DHM. In short, the difference of the notation is simply $(\sigma^\mu)^{\rm DHM}=({\bm 1},\bm\sigma)$, $(\bar{\sigma}^\mu)^{\rm DHM}=({\bm 1},-\bm\sigma)$ with $(+---)$ convention in DHM while $(\sigma^\mu)^{\rm WB}=(-{\bm 1},\bm\sigma)$, $(\bar{\sigma}^\mu)^{\rm WB}=(-{\bm 1},-\bm\sigma)$ with $(-+++)$ convention in WB. We will use the DHM results and its generalization with the WB notation. 

\section{Quantization of spinor fields with time-dependent masses}\label{appB}
We review quantization process of Dirac spinor fields with a time-dependent mass within two component spinor formalism following~\cite{Sou:2021juh} (see also~\cite{Adshead:2015kza}).\footnote{The notation of \cite{Sou:2021juh} is that in DHM~\cite{Dreiner:2008tw}, which differs from ours.}
We consider a free fermionic field with a time-dependent mass
\begin{align}
    \mathcal{L}=\ri\partial_\mu\bar{\psi}_{\dot\alpha}(\bar{\sigma}^\mu)^{\dot\alpha \alpha}\psi_\alpha+\ri\partial_\mu\bar{\tilde{\psi}}_{\dot\alpha}(\bar{\sigma}^\mu)^{\dot\alpha \alpha}\tilde{\psi}_\alpha+m(t)\tilde{\psi}^\alpha\psi_\alpha+\bar{m}(t)\bar{\tilde\psi}_{\dot\alpha}\bar{\psi}^{\dot\alpha},
\end{align}
and note that $\ri\partial_\mu\bar{\psi}_{\dot\alpha}(\bar{\sigma}^\mu)^{\dot\alpha \alpha}\psi_\alpha=-\ri\psi^\alpha(\sigma^\mu)_{\alpha\dot\alpha }\partial_\mu\bar{\psi}^{\dot\alpha}$. 
The canonical conjugate momenta for each spinor field are given by
\begin{align}
    \pi_\psi^\alpha=\frac{\delta \mathcal L}{\delta\dot{\psi}_\alpha}=-\ri\bar{\psi}_{\dot\alpha}(\bar{\sigma}^0)^{\dot\alpha \alpha}, \quad \pi_{\bar\psi\dot\alpha}=\frac{\delta\cal L}{\delta\dot{\bar{\psi}}^{\dot\alpha}}=-\ri\psi^\alpha(\sigma^0)_{\alpha\dot\alpha},
\end{align}
and similarly for $\tilde\psi$. Now the canonical commutation relation reads
\begin{align}
    \left\{\hat{\psi}_\alpha(\bm x),\hat{\pi}^\beta_{\psi}(\bm y)\right\}=\ri\delta^3(\bm x-\bm y)\leftrightarrow \left\{\hat{\psi}_\alpha(\bm x),\hat{\bar{\psi}}_{\dot\alpha}(\bm y)\right\}=-(\sigma^0)_{\alpha\dot\alpha}\delta^3(\bm x -\bm y),
\end{align}
and we omit the relation for $\tilde{\psi}$. In order to expand the spinors in the helicity eigenbasis, we introduce the helicity operator\footnote{$\sigma^0$ or its conjugate expressions are necessary to keep the dotted and undotted indices, but they are essentially an identity matrix. Since the helicity operator consists of a linear combination of the Pauli matrices that form with $\sigma^0\propto {\bm 1}_{2\times2}$ an orthonormal set of $2\times2$ matrices, $\sigma^0$ is only possible matrix to change the indices appropriately while keeping the helicity operator to be $\bm k\cdot \bm\sigma $. In the following, it would be useful to keep in mind that $\sigma_0,\bar\sigma_0\sim {\bm 1}_{2\times2}$.}
\begin{align}
\hat{\bm k}\cdot{\bm\sigma}\to\hat{k}_i(\sigma^0_{\alpha\dot\alpha}(\bar{\sigma}^i)^{\dot\alpha\beta}),\quad (\bar{\sigma}_0)^{\dot\beta\alpha}(\hat{k}_i\sigma^i)_{\alpha\dot\alpha}.
\end{align}
The corresponding eigenspinors of each are given by\
\begin{align}
  & ((\hat{\bm k}\cdot{\bm\sigma}){\bm\xi}_h)_{\alpha}=\hat{k}_i(\sigma^0_{\alpha\dot\alpha}(\bar{\sigma}^i)^{\dot\alpha\beta})\xi_\beta^{h}(\hat{\bm k})=h\xi^{h}_\alpha(\hat{\bm k})\\
 & ({\bm \xi}_h^\dagger(\hat{\bm k}\cdot{\bm\sigma}))_{\dot\alpha} =\xi^\dagger_{h\dot\beta}(\hat{\bm k})(\bar{\sigma}_0)^{\dot\beta\alpha}(\hat{k}_i\sigma^i)_{\alpha\dot\alpha}=h\xi_{h\dot\alpha}^{\dagger}(\hat{\bm k})
\end{align}
where $\hat{k}_i$ is a unit three-momentum vector satisfying $|\hat{\bm k}|^2=1$.\footnote{The corresponding projection operators are
\begin{align}
&P_\pm{}_{\alpha}^\beta=\delta_\alpha^\beta\pm\hat{k}_i(\sigma^0_{\alpha\dot\alpha}(\bar{\sigma}^i)^{\dot\alpha\beta}),\\
&\tilde{P}_\pm{}_{\dot\alpha}^{\dot\beta}=\delta^{\dot\beta}_{\dot\alpha}\pm (\bar{\sigma}_0)^{\dot\beta\alpha}(\hat{k}_i\sigma^i)_{\alpha\dot\alpha}.
\end{align}
} The helicity eigenspinors are normalized as
\begin{align}
    \xi^\dagger_{h\dot\alpha}(\hat{\bm k})\bar{\sigma}_0^{\dot\alpha\beta}\xi_{g\beta}(\hat{\bm k})=\delta_{hg}, \quad \sum_{s=\pm}\xi_{s\alpha}(\hat{\bm k})\xi_{s\dot\beta}^{\dagger}=\sigma_{0\alpha\dot\beta}
\end{align}
(The consistency check: $\xi_{h\alpha}(\hat{\bm k})=(\sigma_0)_{\alpha\dot\beta}\bar{\sigma}_0^{\dot\beta\gamma}\xi_{h\gamma}=(\sum_{s}\xi_{s\alpha}\xi^\dagger_{s\dot\beta})\bar{\sigma}_0^{\dot\beta\gamma}\xi_{h\gamma}=\sum_{s}\xi_{s\alpha}\delta_{sh}=\xi_{h\alpha}$)
Another important relation is
\begin{align}
\xi_{h\dot\beta}^{\dagger}(-\hat{\bm k})\bar{\sigma}_0^{\dot\beta\alpha}\equiv \iota_h(\hat{\bm k})\xi_{h}^\alpha(\hat{\bm k}),\label{flip}
\end{align}
where $\iota_h(\hat{\bm k})$ is a phase factor satisfying $\iota_h(-\hat{\bm k})=-\iota_h(\hat{\bm k})$. 
We formally expand the spinor fields as
\begin{align}
    \hat{\psi}^\alpha(t,\bm x)=&\sum_{h=\pm}\int \frac{d^3\bm k}{(2\pi)^\frac32}\left[\hat{c}_{\bm k,h}\chi_{k,h}(t)e^{\ri\bm k\cdot\bm x}\xi_h^\alpha(\hat{\bm k})+\hat{d}^\dagger_{\bm k,h}\bar{\eta}_{k,h}(t)e^{-\ri{\bm k}\cdot\bm x}\xi^{\dagger}_{h\dot\beta}(\hat{\bm k})\bar\sigma_0^{\dot\beta\alpha}\right],\label{ps}\\
    \hat{\tilde\psi}^\alpha(t,\bm x)=&\sum_{h=\pm}\int \frac{d^3\bm k}{(2\pi)^\frac32}\left[\hat{d}_{\bm k,h}\chi_{k,h}(t)e^{\ri\bm k\cdot\bm x}\xi_h^\alpha(\hat{\bm k})+\hat{c}^\dagger_{\bm k,h}\bar{\eta}_{k,h}(t)e^{-\ri{\bm k}\cdot\bm x}\xi^{\dagger}_{h\dot\beta}(\hat{\bm k})\bar\sigma_0^{\dot\beta\alpha}\right]\label{pst},
\end{align}
where $\chi_h,\eta_h$ are complex scalar functions. The conjugations of them read\footnote{A simple way to compute the conjugate of  $\xi^\dagger_{h\dot\beta}\bar\sigma^{\dot\beta\alpha}_0$ is as follows: Consider $\xi^\dagger_{h\dot\beta}\bar\sigma^{\dot\beta\alpha}_0\Xi_\alpha$ where $\Xi_\alpha$ is a spinor. The conjugation of this product is $\Xi^\dagger\bar{\sigma}_0\xi=\Xi^\dagger_{\dot\alpha}\bar{\sigma}_0^{\dot\alpha\beta}\xi_\beta$, which means that $(\xi^\dagger_{h\dot\beta}\bar\sigma^{\dot\beta\alpha}_0)^\dagger=\bar{\sigma}_0^{\dot\alpha\beta}\xi_\beta$.}
\begin{align}
    \hat{\bar\psi}^{\dot\alpha}(t,\bm x)=&\sum_{h=\pm}\int \frac{d^3\bm k}{(2\pi)^\frac32}\left[\hat{d}_{\bm k,h}\eta_{k,h}(t)e^{\ri\bm k\cdot\bm x}\bar{\sigma}_0^{\dot\alpha\beta}\xi_{h\beta}+\hat{c}^\dagger_{\bm k,h}\bar{\chi}_{k,h}(t)e^{-\ri{\bm k}\cdot\bm x}\xi^{\dagger\dot\alpha}_{h}(\hat{\bm k})\right],\label{psb}\\
   \hat{\bar{\tilde{\psi}}}^{\dot\alpha}(t,\bm x)=&\sum_{h=\pm}\int \frac{d^3\bm k}{(2\pi)^\frac32}\left[\hat{c}_{\bm k,h}\eta_{k,h}(t)e^{\ri\bm k\cdot\bm x}\xi_{h}^\beta(\hat{\bm k})\bar{\sigma}_0^{\dot\alpha\beta}\xi_{h\beta}+\hat{d}^\dagger_{\bm k,h}\bar{\chi}_{k,h}(t)e^{-\ri{\bm k}\cdot\bm x}\xi^{\dagger\dot\alpha}_{h}(\hat{\bm k})\right].\label{pstb}
\end{align}

Note that the pair $\psi,\bar{\tilde{\psi}}$ form a Dirac spinor, and therefore, their wave functions are related to each other.
The normalization condition on the mode functions can be found by requiring $\{\hat{c}_{h,\bm k},\hat{c}_{h',\bm k'}^\dagger\}=\delta_{hh'}\delta^3(\bm k-\bm k')=\{\hat{d}_{h,\bm k},\hat{d}_{h',\bm k'}^\dagger\}$ and the canonical anti-commutation relation, from which we find
\begin{align}
    |\chi_{k,h}(t)|^2+|\eta_{k,h}(t)|^2=1,\label{normalizationcond}
\end{align}
for $\forall t\in\mathbb R$, $\forall k\geq0$ and $h=\pm1$.\footnote{Notice that $-\sigma^0={\bm 1}_{2\times2}$. } The Dirac equation reads
\begin{align}
    &\ri \dot{\chi}_{k,h}(t)+kh\chi_{k,h}(t)+\bar{m}(t)\eta_{k,h}(t)=0,\\
    &\ri \dot{\eta}_{k,h}(t)-kh\eta_{k,h}(t)+m(t)\chi_{k,h}(t)=0,
\end{align}
or equivalently
\begin{align}
    \ri\frac{d}{dt}\left(\begin{array}{c}\chi_{k,h}(t)\\ \eta_{k,h}(t)\end{array}\right)=\left(\begin{array}{cc}
        -kh & -\bar{m}(t) \\
       -m(t)  & kh
    \end{array}\right)\left(\begin{array}{c}\chi_{k,h}(t)\\ \eta_{k,h}(t)\end{array}\right),
\end{align}
and one can easily confirm that these equations ensure \eqref{normalizationcond} to hold as long as it is satisfied at any time $t=t_0$. One may regard this equation as the Schr\"odinger equation for a two level system and instantaneous eigenvalues of the ``Hamiltonian'' are
\begin{align}
    \lambda_{k\pm}(t)=\pm \omega_k(t)\equiv \pm\sqrt{k^2+\mu^2(t)},
\end{align}
where we have defined 
\begin{align}
    m(t)=\mu(t)e^{\ri\phi(t)},
\end{align}
and $\mu(t)$ and $\phi(t)$ are real functions of $t$.
Then we rewrite the ``Schr\"odinger equation'' as
\begin{align}
    \ri\frac{d}{dt}\left(\begin{array}{c}\chi_{k,h}(t)\\ \eta_{k,h}(t)\end{array}\right)=\omega_{k}(t)\left(\begin{array}{cc} \cos\theta_{k,h}(t)& \sin\theta_{k,h}(t) e^{-\ri\phi(t)} \\
       \sin\theta_{k,h}(t) e^{\ri\phi(t)}  & -\cos\theta_{k,h}(t)
    \end{array}\right)\left(\begin{array}{c}\chi_{k,h}(t)\\ \eta_{k,h}(t)\end{array}\right),\label{SchEQ}
\end{align}
where  $\cos\theta_{k,h}(t)\equiv-\frac{kh}{\omega_k(t)}$ and $\sin\theta_{k,h}(t)\equiv -\frac{\mu(t)}{\omega_k(t)}$. The instantaneous eigenvectors of the ``Hamiltonian'' are
\begin{align}
{\bm v}_{k,h}^+(t)\equiv\left(\begin{array}{c} e^{-\ri\phi(t)}\cos\frac{1}{2}\theta_{k,h}(t)\\ \sin\frac{1}{2}\theta_{k,h}(t)
    \end{array}\right),\qquad
{\bm v}_{k,h}^-(t)\equiv \left(\begin{array}{c} -e^{-\ri\phi(t)}\sin\frac{1}{2}\theta_{k,h}(t)\\ \cos\frac{1}{2}\theta_{k,h}(t)
    \end{array}\right) ,
\end{align}
which satisfy ${\bm v}_{k,h}^{\pm\dagger}(t){\bm v}_{k,h}^\pm(t)=1$ and ${\bm v}_{k,h}^{\mp\dagger}(t){\bm v}_{k,h}^\pm(t)=0$.
Taking account of them, we introduce the adiabatic mode functions
\begin{align}
   \left(\begin{array}{c}\chi_{k,h}(t)\\ \eta_{k,h}(t)\end{array}\right)=\alpha_{k,h}(t)e^{-\ri\int^t dt'\omega_k(t')}{\bm v}_{k,h}^+(t)+\beta_{k,h}(t)e^{+\ri\int^t dt'\omega_k(t')}{\bm v}_{k,h}^-(t),\label{Fadexpand}
\end{align}
where we have introduced auxiliary functions $\alpha_{k,h}(t),\beta_{k,h}(t)$ whose evolution equation can be derived from \eqref{SchEQ}. Notice also that the normalization condition $|\chi_{k,h}|^2+|\eta_{k,h}|^2=1$ reads
\begin{align}
    |\alpha_{k,h}(t)|^2+|\beta_{k,h}(t)|^2=1.
\end{align}
The adiabatic vacuum initial condition we would impose is $\alpha_{k,h}(t)\to 1$ and $\beta_{k,h}(t)\to 0$ as $t\to -\infty$ at which the adiabatic condition $\dot{\omega}_{k,h}/\omega_{k,h}^2\to 0$ is satisfied. 

With the formal solution, we are able to define ``time dependent particle number'' in the following way: Substituting the above formal solution to \eqref{ps} with \eqref{flip} yields
\begin{align}
    \hat{\psi}^\alpha(t,\bm x)=&\sum_{h=\pm}\int \frac{d^3k}{(2\pi)^{\frac32}}\nonumber\\
    &\Biggl[\hat{c}_{\bm k,h}e^{+\ri\bm k \cdot\bm x}\xi^\alpha_h(\hat{\bm k})e^{-\ri\phi(t)}\left(\alpha_{k,h}(t)e^{-\ri\int^tdt'\omega_k(t')}\cos\frac12\theta_{k,h}(t)-\beta_{k,h}(t)e^{\ri\int^tdt'\omega_k(t')}\sin\frac12\theta_{k,h}(t)\right)\nonumber\\
    &+\hat{d}^\dagger_{\bm k,h}e^{-\ri{\bm k}\cdot\bm x}\xi^{\dagger}_{h\dot\beta}(\hat{\bm k})\bar\sigma_0^{\dot\beta\alpha}\left(\bar{\alpha}_{k,h}(t)e^{\ri\int^tdt'\omega_k(t')}\sin\frac12\theta_{k,h}(t)+\bar{\beta}_{k,h}(t)e^{-\ri\int^tdt'\omega_k(t')}\cos\frac12\theta_{k,h}(t)\right)\Biggr]\nonumber\\
    =&\sum_{h=\pm}\int \frac{d^3k}{(2\pi)^{\frac32}}\Biggl[e^{-\ri\phi(t)}\cos\frac12\theta_{k,h}(t)e^{\ri\bm k\cdot\bm x}e^{-\ri\int^tdt'\omega_k(t')}\hat{C}_{\bm k,h}(t)\xi^\alpha_h(\hat{\bm k})\nonumber\\
    &\hspace{2.5cm}+\sin\frac12\theta_{k,h}(t)e^{+\ri\bm k\cdot\bm x}e^{+\ri\int^tdt'\omega_k(t')}\hat{D}^\dagger_{\bm k,h}(t)\xi^\dagger_{h\dot\beta}(\hat{\bm k})\bar{\sigma}^{\dot\beta\alpha}_0\Biggr],
\end{align}
where we have defined a new set of creation and annihilation operators
\begin{align}
    &\hat{C}_{\bm k,h}(t)\equiv \alpha_{k,h}(t)\hat{c}_{\bm k,h}+\bar{\beta}_{k,h}(t)e^{\ri\phi(t)}\iota_h(\hat{\bm k})\hat{d}^\dagger_{-\bm k,h},\\
    &\hat{D}^\dagger_{\bm k,h}(t)\equiv \bar{\alpha}_{k,h}(t)\hat{d}^\dagger_{\bm k,h}-\beta_{k,h}(t)e^{-\ri\phi(t)}\iota^*_h(\hat{\bm k})\hat{c}_{-\bm k,h}.
\end{align}
Notice that the adiabatic initial condition $\alpha_k(t)\to1$ and $\beta_k(t)\to 0$ as $t\to-\infty$ implies that $\hat{C}_{\bm k,h}(t)\to\hat{c}_{\bm k,h}$ and $\hat{D}_{\bm k,h}\to \hat{d}_{\bm k,h}$ as expected. Thus, the new set of the creation and annihilation operators is nothing but the Bogoliubov transformation of the past creation and annihilation operators. With the past adiabatic vacuum $\hat{c}_{\bm k,h}|0\rangle_{\rm in}=\hat{d}_{\bm k,h}|0\rangle_{\rm in}=0$, the expectation value of the time-dependent number density $\hat{N}_{\bm k,h}\equiv \hat{C}^\dagger_{\bm k,h}(t)\hat{C}_{\bm k,h}(t)$ and $\hat{\tilde{N}}_{\bm k,h}\equiv \hat{D}^\dagger_{\bm k,h}(t)\hat{D}_{\bm k,h}(t)$ evaluated in the past adiabatic vacuum are
\begin{align}
    {}_{\rm in}\langle 0|\hat{N}_{\bm k,h}(t)|0\rangle_{\rm in}= {}_{\rm in}\langle 0|\hat{\tilde{N}}_{\bm k,h}(t)|0\rangle_{\rm in}=|\beta_{k,h}(t)|^2.
\end{align}
Thus, the number density of the fermionic field produced from the vacuum can be estimated by evaluating $\beta_{k,h}(t)$.

Let us derive the evolution equation of $\alpha_{k,h}(t)$ and $\beta_{k,h}(t)$. Substituting \eqref{Fadexpand} to \eqref{SchEQ}, we find the right-hand-side to be
\begin{align}
   \text{r.h.s. of \eqref{SchEQ}}=\omega_{k}(t)\left(\alpha_{k,h}(t)e^{-\ri\int^t dt'\omega_k(t')}{\bm v}_{k,h}^+(t)-\beta_{k,h}(t)e^{+\ri\int^t dt'\omega_k(t')}{\bm v}_{k,h}^-(t)\right)
\end{align}
whereas the left-hand-side becomes
\begin{align}
    \text{l.h.s. of \eqref{SchEQ}}=&\ri\dot{\alpha}_{k,h}(t)e^{-\ri\int^t dt'\omega_k(t')}{\bm v}_{k,h}^+(t)+\ri\dot{\beta}_{k,h}(t)e^{+\ri\int^t dt'\omega_k(t')}{\bm v}_{k,h}^-(t)\nonumber\\
    &+\ri\alpha_{k,h}(t)e^{-\ri\int^t dt'\omega_k(t')}\dot{\bm v}_{k,h}^+(t)+\ri\beta_{k,h}(t)e^{+\ri\int^t dt'\omega_k(t')}\dot{\bm v}_{k,h}^+(t)\nonumber\\
    &+\omega_{k}(t)\left(\alpha_{k,h}(t)e^{-\ri\int^t dt'\omega_k(t')}{\bm v}_{k,h}^+(t)-\beta_{k,h}(t)e^{+\ri\int^t dt'\omega_k(t')}{\bm v}_{k,h}^-(t)\right).
\end{align}
Thus, comparison of the both side and multiplication of either $\langle E^\pm_{k,h}(t)|$ give us
\begin{align}
   & \dot{\alpha}_{k,h}(t)=-\alpha_{k,h}(t){\bm v}_{k,h}^{+\dagger}(t)\dot{\bm v}_{k,h}^+(t)-\beta_{k,h}(t)e^{+2\ri\int^tdt'\omega_k(t')}{\bm v}_{k,h}^{+\dagger}(t)\dot{\bm v}_{k,h}^-(t),\\
   &\dot{\beta}_{k,h}(t)=-\alpha_{k,h}(t)e^{-2\ri\int^tdt'\omega_k(t')}{\bm v}_{k,h}^{-\dagger}(t)\dot{\bm v}_{k,h}^+(t)-\beta_{k,h}(t){\bm v}_{k,h}^{-\dagger}(t)\dot{\bm v}_{k,h}^-(t).
\end{align}
Notice that 
\begin{align}
    \dot{\bm v}_{k,h}^+(t)=&\frac12\dot\theta_{k,h}(t){\bm v}_{k,h}^-(t)-\ri\dot\phi(t)\left(\begin{array}{c}e^{-\ri\phi(t)}\cos\frac12\theta_{k,h}(t)\\0\end{array}\right),\nonumber\\
    \dot{\bm v}_{k,h}^-(t)=&-\frac12\dot\theta_{k,h}(t){\bm v}_{k,h}^+(t)+\ri\dot\phi(t)\left(\begin{array}{c}e^{-\ri\phi(t)}\sin\frac12\theta_{k,h}(t)\\0\end{array}\right),
\end{align}
and therefore,
\begin{align}
     & \dot{\alpha}_{k,h}(t)=\frac{\ri}{2}\alpha_{k,h}(t)\dot\phi(1+\cos\theta_{k,h}(t))+\frac12\beta_{k,h}e^{+2\ri\int^tdt'\omega_k(t')}\left(\dot\theta_{k,h}(t)-\ri\dot\phi(t)\sin\theta_{k,h}(t)\right),\nonumber\\
     &\dot{\beta}_{k,h}(t)=-\frac12\alpha_{k,h}(t)e^{-2\ri\int^tdt'\omega_k(t')}\left(\dot\theta_{k,h}(t)+\ri\dot\phi(t)\sin\theta_{k,h}(t)\right)+\frac{\ri}{2}\dot\phi(t)\beta_{k,h}(1-\cos\theta_{k,h}(t)).
\end{align}
By the redefinition $\alpha_{k,h}=e^{\ri\int^tdt'\omega_k(t')}\tilde{\alpha}_{k,h}$ and $\beta_{k,h}=e^{-\ri\int^tdt'\omega_k(t')}\tilde{\beta}_{k,h}$ and noting $\dot\alpha_{k,h}=(\ri\omega_k\tilde{\alpha}_{k,h}+\dot{\tilde{\alpha}}_{k,h})e^{\ri\int^tdt'\omega_k(t')}$ and $\dot\beta_{k,h}=(-\ri\omega_k\tilde{\beta}_{k,h}+\dot{\tilde{\beta}}_{k,h})e^{-\ri\int^tdt'\omega_k(t')}$, we obtain
\begin{align}
&\dot{\tilde{\alpha}}_{k,h}=-\ri\omega_k\tilde{\alpha}_{k,h}+\frac{\ri}{2}\tilde{\alpha}_{k,h}\dot\phi(1+\cos\theta_{k,h})+\frac12\tilde{\beta}_{k,h}\left(\dot\theta_{k,h}-\ri\dot\phi\sin\theta_{k,h}\right),\nonumber\\
&\dot{\tilde{\beta}}_{k,h}(t)=\ri\omega_k\tilde{\beta}_{k,h}-\frac12\tilde{\alpha}_{k,h}\left(\dot\theta_{k,h}+\ri\dot\phi\sin\theta_{k,h}\right)+\frac{\ri}{2}\dot\phi\tilde{\beta}_{k,h}(1-\cos\theta_{k,h}).\label{tildeabeq}
\end{align}
One can check that $\frac{d}{dt}(|\tilde{\alpha}_{k,h}|^2+|\tilde{\beta}_{k,h}|^2)=0$ by the above equations, which proves that the normalization condition holds at an arbitrary time.

Solving \eqref{tildeabeq} suffices for evaluation of the particle number density as $|\tilde\beta_{k,h}(t)|^2=|\beta_{k,h}(t)|^2$ and therefore, by solving these equations either analytically or numerically one can evaluate the produced number density. These equations are practically more tractable as there appear no integrated phase factor $e^{\pm\ri\int^tdt'\omega_k(t')}$, which may be reintroduced after evaluating $(\tilde{\alpha}_{k,h},\tilde{\beta}_{k,h})$ if necessary. 

We emphasize that the ``time-dependent Bogoliubov coefficients'' $\alpha_{k,h}(t)$ and $\beta_{k,h}(t)$ introduced as coefficients of adiabatic mode functions are not unique. Indeed the adiabatic mode function we have used here is the lowest order in adiabatic expansion, and one could choose higher-order adiabatic mode functions with which the behavior of ``time-dependent Bogoliubov coefficients'' differs from ours. See~\cite{Berry:1990,Lim:1991,Dabrowski:2014ica,Dabrowski:2016tsx,Yamada:2021kqw,Ilderton:2021zej} for details. Despite the smoothing criteria introduced by Berry~\cite{Berry:1990,Lim:1991}, there is no unique way to define the notion of particles or equivalently vacuum states at the time when the time-dependent background field is yet turned on. Nevertheless, the number density of asymptotic region without time-dependent background fields turn out to be independent of the choice of the vacuum state at intermediate time, and therefore, can have a definite physical meaning.

\subsection{Quantization of a complex scalar field with time-dependent mass}
For completeness, we also discuss quantization of a complex scalar field with a time-dependent mass in almost the same way as a Dirac field. The system we consider consists of a complex scalar $\Phi$ with a time dependent mass $m^2(t)$,
\begin{align}
    S=-\int d^4x\left[|\partial\Phi|^2+m^2(t)|\Phi|^2\right].
\end{align}
We expand a quantized field $\hat\Phi$ as
\begin{align}
    \hat{\Phi}(t,\bm x)=\int \frac{d^3k}{(2\pi)^\frac32}\left[\hat{a}_{\bm k}f_k(t)e^{\ri\bm k\cdot\bm x}+\hat{b}^\dagger_{\bm k}\bar{f}_{k}(t)e^{-\ri\bm k\cdot\bm x}\right]
\end{align}
where we have introduced creation and annihilation operators that obey the commutation relation $[\hat{a}_{\bm k},\hat{a}^\dagger_{\bm k'}]=\delta^3(\bm k-\bm k')=[\hat{b}_{\bm k},\hat{b}^\dagger_{\bm k'}]$ and the mode function $f_k(t)$ satisfies
\begin{align}
\ddot{f}_k(t)+\left(k^2+m^2(t)\right)f_k(t)=0
\end{align}
following from the equation of motion of $\Phi$. As is the case of the Dirac field, one can write a formal solution of the mode equation as
\begin{align}
    f_{k}(t)=\frac{1}{\sqrt{2\omega_k(t)}}\left[\gamma_k(t)e^{-\ri\int^tdt'\omega_k(t')}+\delta_k(t)e^{\ri\int^tdt'\omega_k(t')}\right],
\end{align}
where $\omega_k(t)\equiv \sqrt{k^2+m^2(t)}$ and the auxiliary function $\gamma_k(t),\delta_k(t)$ are normalized as $|\gamma_k(t)|^2-|\delta_k(t)|^2=1$.\footnote{Notice the difference of the sign of the second term from the fermionic case.} We have chosen the auxiliary functions to satisfy
\begin{align}
    \frac{d}{dt}\left(\begin{array}{c}\gamma_k(t)\\ \delta_k(t)
    \end{array}\right)=\left(\begin{array}{cc}0&\frac{\dot{\omega}_k(t)}{2\omega_k(t)}e^{2\ri \int^t dt'\omega_k(t')}\\
    \frac{\dot{\omega}_k(t)}{2\omega_k(t)}e^{-2\ri \int^t dt'\omega_k(t')}&0\end{array}\right)\left(\begin{array}{c}\gamma_k(t)\\ \delta_k(t)
    \end{array}\right).
\end{align}
We emphasize that like the Dirac field case, there are ambiguities of the choice of the base function, and we have chosen the lowest order adiabatic solution. Furthermore, at least for bosonic cases, there appears an ambiguity in the choice of the evolution equation of $\gamma_k(t),\delta_k(t)$. Note however that such ambiguities appear only during the time interval on which the time-dependent background fields are turned on, and the physical quantities such as particle number density in the asymptotic region without time-dependent backgrounds are not affected by such ambiguities in the choice of the vacuum state at intermediate time (see \cite{Berry:1990,Lim:1991,Dabrowski:2014ica,Dabrowski:2016tsx,Yamada:2021kqw,Ilderton:2021zej} for details). As the spinor case, it would be more practically useful to introduce variables $\tilde{\gamma}_k(t)\equiv\gamma_k(t)e^{-\ri \int^t dt'\omega_k(t')}$ and $\tilde{\delta}_k(t)\equiv\delta_k(t)e^{\ri \int^t dt'\omega_k(t')}$ whose time evolution equations are 
\begin{align}
    \frac{d}{dt}\left(\begin{array}{c}\tilde{\gamma}_k(t)\\ \tilde{\delta}_k(t)
    \end{array}\right)=\left(\begin{array}{cc}\ri\omega_k(t)&\frac{\dot{\omega}_k(t)}{2\omega_k(t)}\\
    \frac{\dot{\omega}_k(t)}{2\omega_k(t)}&-\ri\omega_k(t)\end{array}\right)\left(\begin{array}{c}\tilde{\gamma}_k(t)\\ \tilde{\delta}_k(t)
    \end{array}\right).
\end{align}
Once again, the advantage of these variables is that the phase factor $e^{\pm\ri \int^t dt'\omega_k(t')}$, which are technically difficult to compute in general, disappears from the evolution equation.

By the same reasons as the Dirac field case, $(\gamma_k(t),\delta_k(t))$ can be understood as the ``time-dependent Bogoliubov coefficients'' and the particle number density produced from the past adiabatic vacuum is expressed as
\begin{align}
    {}_{\rm in}\langle0|\hat{N}_{\bm k}(t)|0\rangle_{\rm in}=|\delta_k(t)|^2=|\tilde{\delta}_k(t)|^2,
\end{align}
where $\hat{N}_{\bm k}(t)=\hat{A}^\dagger(t)\hat{A}(t)$ and $\hat{A}(t)\equiv \gamma_k(t)\hat{a}_{\bm k}+\delta^*_k(t)\hat{b}^\dagger_{-\bm k}$ which corresponds to the annihilation operator of the instantaneous vacuum defined at the time $t$. Here we have assumed the adiabatic vacuum mode function $\gamma_k(t)\to1,\ \delta_k(t)\to0$ as $t\to-\infty$ and used the in-vacuum state satisfies $\hat{a}_{\bm k}|0\rangle_{\rm in}=\hat{b}_{\bm k}|0\rangle_{\rm in}=0$ for $\forall \bm k\in {\mathbb R}^3$.

\section{External potential approach to KK expansion of matter coupled to the electric field on magnetized torus}\label{externalpotential}
In this section, we show that treating time-dependent gauge potential as the external potential results in an inconvenient form of effective action. In Sec.~\ref{TDWL}, we instead include the time-dependent gauge potential as time-dependent Wilson line and solve the mode function exactly. We will see that the result of this section would consistent with the time-dependent Wilson line approach.

In the following, we will expand all charged superfields with the wave functions {\it without} time-dependent Wilson lines. Similarly to the case in Sec.~\ref{TDWL}, using the algebra of Dirac operators as well as KK mode expansion, the superpotential reads
\begin{align}
   &\frac{1}{\pi R}\int d^2z\sqrt{g_2}\left[\int d^2\theta \tilde{Q}\left(\partial-\frac{N \bar{z}}{2}\right)Q+\int d^2\bar{\theta}\bar{\tilde Q}\left(\bar{\partial}-\frac{Nz}{2}\right)\bar{Q}\right]\nonumber\\
   &=\frac{\sqrt{\pi N}}{\pi R}\int d^2z\sqrt{g_2}\left[\int d^2\theta \tilde{Q} (\hat{a}^\dagger Q)+\int d^2\bar{\theta}(\hat{a}\bar{\tilde{Q}})\bar{Q}\right]\nonumber\\
   &=\sum_{n\geq0,j}\frac{\sqrt{\pi N}}{\pi R}\left[\int d^2\theta \sqrt{n+1}\tilde{Q}_{n+1,j} Q_{n,j}+\int d^2\bar{\theta}\sqrt{n}\bar{\tilde Q}_{n,j}\bar{Q}_{n-1,j}\right]\nonumber\\
   &=\sum_{n\geq0,j}\sqrt{\frac{4\pi N(n+1)}{\mathcal{A}}}\left[\int d^2\theta \tilde{Q}_{n+1,j} Q_{n,j}+\int d^2\bar{\theta}\bar{\tilde Q}_{n+1,j}\bar{Q}_{n,j}\right],
\end{align}
where $\mathcal{A}=(2\pi R)^2$ is the area of the torus and we have shifted the KK number of the second term in the third equality.
One can perform the integral over the compact space for the electric potential term and finally finds the hypermultiplet action to be (after reintroducing $\zeta$-terms)
\begin{align}
    &\int d^6X\sqrt{G}\left[\int d^4\theta(\tilde{Q} e^{qV}\bar{\tilde{Q}}+\bar{Q}e^{-qV}Q)+\left\{\int d^2\theta \frac{1}{\pi R}\tilde{Q}\left(\partial-\frac{q}{\sqrt2}\phi\right)Q+{\rm h.c.}\right\}\right]\nonumber\\
   =& \int d^4x \sum_{n\geq0} \Biggl[(\tilde{Q}_{n,j}e^{qg_{\rm 4D}V_0}\bar{\tilde{Q}}_{n,j}+\bar{Q}_{n,j}e^{-qg_{\rm 4D}V_0}Q_{n,j})\nonumber\\
   &\qquad+\left\{\int d^2\theta \left(M_{n}\tilde{Q}_{n+1,j}Q_{n,j}-\frac{\zeta(t)}{\sqrt2\pi R}\tilde{Q}_{n,j}Q_{n,j}\right)+{\rm h.c.}\right\}\Biggr],
\end{align}
where $M_n\equiv\sqrt{\frac{4\pi N(n+1)}{\mathcal{A}}}$ and we have left the coupling between the zero mode of $V$ denoted by $V_0$ and the hypermultiplet. Since supersymmetry is spontaneously broken by the magnetic flux or more specifically by the flux-induced Feyet-Iliopoulos term, we need to consider mass matrices of bosonic and fermionic components separately. In components, the terms containing auxiliary fields are
\begin{align}
  S_{\rm F,D}=&\int d^4x\Biggl[\frac12 D_{0,0}^2+\frac12 D_{0,0}\left(\frac{4N}{R q g}+\sum_{n}\{qg_{\rm 4D}|\tilde{Q}_{n,j}|^2-qg_{\rm 4D}|Q_{n,j}|^2\}\right)+|\tilde{F}_{n,j}|^2+|F_{n,j}|^2\nonumber\\
 & +\left(\tilde{F}_{n,j}\left(M_{n-1}Q_{n-1,j}-\frac{gq\zeta}{\sqrt2\pi R}Q_{n,j}\right)+F_{n,j}\left(M_{n}\tilde{Q}_{n+1,j}-\frac{gq\zeta}{\sqrt2\pi R}\tilde{Q}_{n,j}\right)+{\rm h.c.}\right)\Biggr]\nonumber\\
\to &-\frac18\left(\frac{8\pi N}{{\mathcal A} q g_{\rm 4D}}+\sum_{n}\{qg_{\rm 4D}|\tilde{Q}_{n,j}|^2-qg_{\rm 4D}|Q_{n,j}|^2\}\right)^2\nonumber\\
&-\sum_{n\geq0}\left[\left|M_{n-1}Q_{n-1,j}-\frac{gq\zeta}{\sqrt2\pi R}Q_{n,j}\right|^2+\left|M_{n}\tilde{Q}_{n+1,j}-\frac{gq\zeta}{\sqrt2\pi R}\tilde{Q}_{n,j}\right|^2\right],
\end{align}
where the arrow indicates the on-shell action with respect to auxiliary fields and it is understood that $\tilde{Q}_{-1}=0$. Note that for $\zeta=0$ we find the mass terms of the hyper-scalars as
\begin{align}
    \mathcal{L}=&\sum_{n\geq0}\left[-\frac{2\pi N}{\mathcal{A}}(|\tilde{Q}_{n,j}|^2-|Q_{n,j}|^2)-\frac{4\pi Nn}{\mathcal A}|Q_{n-1,j}|^2-\frac{4\pi N(n+1)}{\mathcal A}|\tilde{Q}_{n+1,j}|^2\right]\nonumber\\
    =&-\sum_{n\geq0}\frac{2\pi N(2n+1)}{\mathcal A}(|\tilde{Q}_{n,j}|^2+|Q_{n,j}|^2),
\end{align}
which means that the mass eigenvalues of the hyper-scalars are $m_n^2=\frac{2\pi N(2n+1)}{\mathcal A}$. Note that there appears a positive cosmological constant term due to the magnetic flux energy density, which we will do not discuss further. On the basis of the above case, let us turn to the case with electric field. Turning on the electric field yields the following scalar mass terms:
\begin{align}
    \mathcal{L}_{\rm s.m.}=&-\sum_{n\geq 0}\left[\frac{2\pi N}{\mathcal{A}}(|\tilde{Q}_{n,j}|^2-|Q_{n,j}|^2)+\left|M_{n-1}Q_{n-1,j}-\frac{gq\zeta}{\sqrt2\pi R}Q_{n,j}\right|^2+\left|M_{n}\tilde{Q}_{n+1,j}-\frac{gq\zeta}{\sqrt2\pi R}\tilde{Q}_{n,j}\right|^2\right]\nonumber\\
    =&-\sum_{n\geq0}\frac{2\pi N(2n+1)}{\mathcal A}(|\tilde{Q}_{n,j}|^2+|Q_{n,j}|^2)-\sum_{n\geq0}2(g_{\rm 4D}q)^2|\zeta|^2(|Q_{n,j}|^2+|\tilde{Q}_{n,j}|^2)\nonumber\\
    &+\sum_{n\geq 0}2 \sqrt{\frac{2\pi N(n+1)}{\mathcal A}}g_{\rm 4D}q\left(\bar{\zeta}Q_{n,j}\bar{Q}_{n+1,j}+\zeta\tilde{Q}_{n,j}\bar{\tilde{Q}}_{n+1,j}+{\rm h.c.}\right)\nonumber\\
    =&-\sum_{n\geq0}\Biggl[-\frac{2\pi N}{\mathcal{A}}|Q_{n,j}|^2+\left|\sqrt{\frac{4\pi N(n+1)}{\mathcal A}}Q_{n,j}-\sqrt2 (g_{\rm 4D}q)\zeta Q_{n+1,j}\right|^2\nonumber\\
    &\qquad \quad-\frac{2\pi N}{\mathcal{A}}|\tilde{Q}_{n,j}|^2+\left|\sqrt{\frac{4\pi N(n+1)}{\mathcal A}}\tilde{Q}_{n,j}-\sqrt2 (g_{\rm 4D}q)\bar{\zeta} \tilde{Q}_{n+1,j}\right|^2\Biggr].
\end{align}
Notice that the mass terms are independent of the degeneracy index $j$ and therefore, we will omit it in the following. Defining ${\bm Q}=(Q_0,Q_1,\cdots)^{\rm t}$ and similarly for $\tilde{\bm Q}$ where $\rm t$ denotes transpose, we may write the mass terms as $-{\bm Q}^\dagger \mathcal{M}^2_B{\bm Q}-\tilde{\bm Q}^\dagger (\mathcal{M}_B^2)^{\rm t}\tilde{\bm Q}$ where
\begin{align}
     \mathcal{M}_B^2\equiv\left(\begin{array}{ccccc}
        M_0^2-m_0^2+|\tilde\zeta|^2&-\tilde{\zeta}M_0 &0&0&\cdots \\
       -\bar{\tilde{\zeta}}M_0& M_1^2-m_0^2+|\tilde\zeta|^2&-\tilde{\zeta}M_1&0&\cdots\\
       0&-\bar{\tilde{\zeta}}M_1&M_2^2-m_0^2+|\tilde\zeta|^2&-\tilde{\zeta}M_2&\cdots\\
       0&0&-\bar{\tilde{\zeta}}M_2&M_3^2-m_0^2+|\tilde\zeta|^2&\cdots\\
    \vdots&\vdots&\vdots&\vdots&\ddots
    \end{array}\right),
\end{align}
where $m_0^2\equiv \frac{2\pi N}{\mathcal A}$ and $\tilde\zeta\equiv \sqrt2qg_{\rm 4D}\zeta$. Thus, we have found the mass matrix to be a tridiagonal form, which shows that the electric field relates nearest neighbor modes. Notice also that when $\zeta$ is turned off, the mass matrix is diagonal.

Next, we consider fermion mass terms, which are not affected by D-term contributions associated with a flux-induced FI term. The superpotential yields the fermionic mass terms
\begin{align}
  \mathcal{L}_{FM}=\sum_{n\geq 0}\left[ M_n\tilde{\psi}_{n+1,j}\psi_{n,j}-\sqrt2 qg_{\rm 4D}\zeta\tilde{\psi}_{n,j}\psi_{n,j}+{\rm h.c.}\right].
\end{align}
Notice that $\tilde{\psi}_{0,j}$ does not appear in the first term since it is the chiral zero mode in the magnetized background. Nevertheless, the electric part yields an effective time-dependent mass to the zero mode. The mass term can be written as $\tilde{\bm\psi}^{\rm t}\mathcal{M}_F{\bm\psi}$ where ${\bm \psi}=(\psi_0,\psi_1,\cdots)^{\rm t}$ and
\begin{align}
    \mathcal{M}_F=\left(\begin{array}{ccccc}
        -\tilde\zeta&0 &0&0&\cdots \\
       M_0& -\tilde\zeta&0&0&\cdots\\
       0&M_1&-\tilde\zeta&0&\cdots\\
       0&0&M_2&-\tilde\zeta&\cdots\\
   \vdots&\vdots&\vdots&\vdots&\ddots
    \end{array}\right).
\end{align}
 The squared mass matrix has the tridiagonal structure as
\begin{align}
    \mathcal{M}_F^\dagger\mathcal{M}_F=\left(\begin{array}{ccccc}
       M_0^2+ |\tilde\zeta|^2&-\tilde{\zeta} M_0 &0&0&\cdots \\
       -\bar{\tilde\zeta}M_0&M_1^2+|\tilde\zeta|^2&-\tilde\zeta M_1&0&\cdots\\
       0&-\bar{\tilde\zeta} M_1&M_2^2+|\tilde\zeta|^2&-\tilde\zeta M_2&\cdots\\
       0&0&-\bar{\tilde\zeta} M_2&M_3^2+|\tilde\zeta|^2&\cdots\\
    \vdots&\vdots&\vdots&\vdots&\ddots
    \end{array}\right)
\end{align}
whereas
\begin{align}
\mathcal{M}_F\mathcal{M}_F^\dagger=\left(\begin{array}{ccccc}
        |\tilde\zeta|^2&-\bar{\tilde\zeta} M_0 &0&0&\cdots \\
       -\tilde\zeta M_0&M_0^2+|\tilde\zeta|^2 &-\bar{\tilde\zeta} M_1&0&\cdots\\
       0&-\tilde\zeta M_1&M_1^2+|\tilde\zeta|^2&-\bar{\tilde\zeta} M_2&\cdots\\
       0&0&-\tilde\zeta M_2&M_2^2+|\tilde\zeta|^2&\cdots\\
    \vdots&\vdots&\vdots&\vdots&\ddots
    \end{array}\right).
\end{align}
Notice that when $\zeta=0$, each matrix is diagonal and the eigenvalues are either $\{M_n^2\}$ or $\{0,M_n^2\}$, respectively. Thus, we find the mass matrix structure similar to the case of hyper-scalars. Note that since there is a non-vanishing FI term, bosonic and fermionic masses cannot be degenerate. 

Let us discuss the mass eigenvalues. Despite explicit bosonic and fermionic mass matrices, we could not find the eigenvalues of the infinite dimensional matrices. Physically, we would not expect all the KK modes to be excited particularly for small $\zeta$. Therefore, we have considered the $m\times m$ sub-matrices ($m=2,3,4$) that consists of the first $n$ row and column. We have checked the eigenvalues of the sub-matrices (with Mathematica) in the weak field limit $\zeta\ll m_0\sim M_0$. A surprising property we find is that up to the quadratic order in $\zeta$, only the $m$-th eigenvalue $\mu_m$ receives a correction $\mathcal{O}(\zeta^2)$. We also find that only the even powers of $\zeta$ appear up to 6th order. From these observations, we expect that in the presence of the magnetic fluxes, the electric field does not change the KK mass eigenvalue. Such a property is in contrast  to the case without magnetic fluxes in which the gauge potential linearly changes the KK mass in the weak field limit, from which we anticipate KK modes that become light due to the electric force deceleration. Notice however that even if mass eigenvalues are not deformed by the time-dependent gauge potential, the diagonalizing matrix for mass terms is in general time-dependent. Therefore, the instantaneous mass eigenbasis acquire additional terms associated with time dependence. For instance, suppose the kinetic basis $\phi_n$ is related to the mass eigenbasis $\varphi_n$ as $\phi_n=O_{nm}(t)\varphi_m$ where $O_{nm}$ the orthogonal matrix. Then, the kinetic term of original basis $\dot{\bar{\phi}}_n\dot{\phi}_n= (\dot{\bar{\varphi}}_m O^{\rm t}_{mn}+\dot{O}^{\rm t}_{mn}\bar{\varphi}_m)(\dot{\varphi}_p O_{np}+\dot{O}_{np}\varphi_p)=\dot{\bar{\varphi}}_m\dot{\varphi}_m+\mathcal{O}(\dot\zeta)$ where $\mathcal{O}(\dot\zeta)$ denotes the terms vanishing in the limit $\zeta\to {\rm const.}$. Therefore, the effects of the time-dependent gauge potential appear through such terms. In other words, the time-dependent gauge potential ``rotates'' the KK mass eigenbasis. 

The expectation that the KK mass is independent of $\zeta$ can actually be explained as follows: The above derivation does not rely on the fact that $\zeta$ is time-dependent, that is, we could find the same mass matrices even when $\zeta$ is constant. However, in such a case, we are able to find a non-perturbative eigenfunction for a Dirac operator $\mathcal{D}_\zeta=\partial-\frac{\pi N\bar{z}}{2}-\frac{qg}{\sqrt2}\zeta$ where $\zeta$ is a constant Wilson line. The presence of Wilson line does not change the algebraic properties of Dirac operators, and therefore, the eigenvalues of the Dirac operator is independent of the constant Wilson line. This implies that the eigenvalues of the infinite dimensional matrices are the same as that without $\zeta(t)$, which is consistent with the above observation. Indeed, the approach in Sec.~\ref{TDWL} clarifies such expectations that the eigenvalues of Dirac/Laplace operators are independent of $\zeta(t)$ but there appears effective terms mixing the different KK levels as expected above.

\section{Semi-realistic models from 10D super Yang-Mills model on tori}\label{10Dreview}
In order for this work to be self-contained, we review the construction of semi-realistic models within 10D super Yang-Mills theory compactified on tori~\cite{Cremades:2004wa,Abe:2012fj,Abe:2012ya}. In this model, the metric is taken as $ds^2=\eta_{\mu\nu}dx^\mu dx^\nu+\sum_{i=1}^3(2\pi R_i)^2dz^id\bar{z}^i$ where $z^i=\frac12 (y^{2+2i}+\tau_iy^{3+2i})$ and $\tau_i$ $(i=1,2,3)$ denote complex structure moduli. We consider $U(N)$ gauge theory and the gauge potential along compact directions become 4D complex scalars
\begin{align}
     {A}^{ab}_i=-\frac{1}{{\rm Im}\tau_i}(\tau_i^*A^{ab}_{2+2i}-A^{ab}_{3+2i}),
 \end{align}
 where $a,b$ are $U(N)$ gauge group indices. Thus, the 4D bosonic degrees of freedom consist of a vector and three complex scalars $(A^{ab}_{\mu},A_i^{ab})$ for each matrix element $(ab)$, whereas the 4D fermionic degrees of freedom originate from a 10D Majorana-Weyl gaugino, which can be decomposed into four 4D Weyl spinors $(\lambda_{+++}^{ab},\lambda_{+--}^{ab},\lambda_{-+-}^{ab}, \lambda_{--+}^{ab})$ and the subscripts denote the chirality on each torus.\footnote{We have chosen the 10D Majorana-Weyl spinor to be even with respect to 10D chirality and therefore the total chirality of each 4D Weyl spinor should be even as well.} These components can be repackaged into vector and chiral superfields
 \begin{align}
     {V}^{ab}(x^\mu,z^i,\bar{z}^i,\theta,\bar\theta)=&\cdots -\theta\sigma^\mu\bar\theta { A}^{ab}_\mu+\ri\bar\theta^2\theta{\lambda}^{ab}_0-\ri\theta^2\bar\theta\bar{\lambda}^{ab}_0+\frac12 {D}^{ab},\\
     {\phi}^{ab}_i(x^\mu,z^i,\bar{z}^i,\theta,\bar\theta)=&\frac{1}{\sqrt2}{ A}^{ab}_i+\sqrt{2}\theta{\lambda}^{ab}_i+\theta{F}^{ab}_i+\cdots,
 \end{align}
 where we have introduced F- and D-terms $F^{ab}_i$ and $D^{ab}$ and the ellipses in a vector superfield denote the components that vanish in Wess-Zumino gauge, while that in a chiral superfield denote the terms that vanish in chiral coordinates. Note that the component fields are yet 10D fields dependent on the coodinates on tori. In Wess-Zumino gauge, the 10D super Yang-Mills action in magnetic flux background is given by
\begin{align}
    K=&\frac{1}{g^2}\sum_{a,b}\sum_{i=1}^3\frac{1}{(\pi R_i)^2}\Biggl[\bar{\phi}^{ab}_{\bar i}{\phi}^{ab}_i+2f_{\bar{i}i}^{a}V^{aa}-\sqrt{2}(\bar{ D}^{\dagger ba}_i\bar{\phi}^{ba}_{\bar i}+{D}^{\dagger ab}_{\bar i}{\phi}^{ab}_i){V}^{ba}+([\bar{\bm\phi}_{\bar i},{\bm\phi}_i])^{ab}{V}^{ba}\nonumber\\
    &-\left({D}^{\dagger ab}_{\bar i}{V}^{ab}-\frac{1}{\sqrt2}([\bar{\bm\phi}_{\bar i},{\bm V}])^{ab}\right)\left(D^{ba}_i{V}^{ba}-\frac{1}{\sqrt2}([{\bm\phi}_i,{\bm V}])^{ba}\right)\Biggr],\\
    W=&\sum_{a,b}\frac{1}{g^2}\frac{1}{2 \prod_{l=1}^3(\pi R_l)}\epsilon^{ijk}\left[\phi_{i}^{ab}D_j^{ba}\phi_k^{ba}-\frac{1}{3\sqrt2}\phi_i^{ab}([{\bm\phi_j,\bm\phi_k}])^{ba}\right],
\end{align}
where $f_{\bar{i}j}^a\equiv \sqrt{2}\partial_j\langle\bar{\phi}^{aa}_{i}\rangle=\sqrt2\bar{\partial}_{\bar i}\langle\phi_j^{aa}\rangle$ and the covariant derivatives are given by $D^{ab}_i=\partial_i-\frac{1}{\sqrt2}(\langle\phi_i^{aa}\rangle-\langle\phi_i^{bb}\rangle)$, $\bar{D}_{\bar i}^{ab}=-\bar{\partial}_{\bar i}-(\langle\bar{\phi}_i^{aa}\rangle-\langle\bar\phi^{bb}_i\rangle)$ and the bold letters denote $U(N)$ matrices. 

Let us introduce Abelian magnetic and electric backgrounds. We here consider the following Abelian magnetic fluxes within $U(N)$ theory:
\begin{equation}
    {\bm F}_{z^i\bar{z}^i}dz^i\wedge d\bar{z}^i=\frac{\ri \pi {\bm M}^{(i)}}{{\rm Im}\tau}dz^i\wedge d\bar{z}^i
\end{equation}
where no sum is taken over $i$, the constant fluxes in a $U(N)$ matrix form are
\begin{equation}
    {\bm M}^{(i)}=\left(\begin{array}{cccc}
        M^{(i)}_1{\bm 1 }_{N_1\times N_1}&{}&{}&{}\\
        {} & M^{(i)}_2{\bm 1}_{N_2\times N_2}&{}&{}\\
        {}&{}&M^{(i)}_3{\bm 1}_{N_3\times N_3}&{}\\
        {}&{}&{}&\ddots\\
    \end{array}\right),\label{genAflux}
\end{equation}
and $\sum_{A=1}^{\tilde N}N_A=N\quad (A=1,2,\cdots,\tilde{N})$ and $M_1^{(i)}\in {\mathbb Z}$ from the Dirac quantization condition.\footnote{We have assumed $M_A\neq M_B$ for $A\neq B$ without loss of generality.} This flux realizes $U(N)\to\prod_{A=1}^{\tilde{N}}U(N_A)$. In general, there can be (time-dependent) Wilson lines that break the gauge groups into smaller ones as in the model that we will discuss later.  

As we have discussed in Sec.~\ref{TDWL}, due to the KK level structure, the KK modes would persist to be heavy even if electric field exist for a sufficiently long time unlike the case of purely electric field cases and therefore not be produced non-perturbatively as long as the electric field strength is smaller than the KK scale. Therefore, we will neglect KK modes and focus on the 4D effective theory including zero modes only. Then, one can drop the terms containing the covariant derivatives along compact directions. We formally expand matter superfield ${\bm\phi}_i$ as
\begin{align}
    \bm\phi_i(x^\mu,z^i,\bar{z}^i,\theta)=\sum_{\bm n}\prod_{j=1}^3 f^{(j),n_j}_i(z^j,\bar{z}^j){\bm\phi}_i^{\bm n}(x^\mu,\theta),
\end{align}
where $\bm n=(n_1,n_2,n_3)$ is the set of KK numbers on each torus and $f_{i}^{(j),n_j}(z^j,\bar{z}^j)$ denotes the $n_j$-th mode function on $j$-th torus. In particular, the zero mode function $n_j=0$ satisfies the eigenequation
\begin{align}
   & \left[\bar{\partial}_{\bar{j}}+\frac{\pi}{2{\rm Im}\tau_i}\left(M_{AB}^{(i)}z^i+\zeta^i_{AB}(t)\right)\right]f^{(j),n_j=0}_{i,AB}=0 \quad \text{for }i= j\nonumber\\
  &  \left[\partial_{j}-\frac{\pi}{2{\rm Im}\tau_i}\left(M_{AB}^{(i)}\bar{z}^i+\bar{\zeta}^i_{AB}(t)\right)\right]f^{(j),n_j=0}_{i,AB}=0 \quad \text{for }i\neq j
\end{align}
where $M_{AB}^{(i)}\equiv M_{A}^{(i)}-M_{B}^{(i)}$ for $A\neq B$ ($A,B=1,\cdots \tilde{N}$). We note that the Wilson lines are taken to be time-dependent explicitly but we will find that it does not change the known results except for the fact that the resultant Yukawa couplings become time-dependent. The solutions can be summarized as follows: For $i=j$ sector,
\begin{align}
    f_{i,AB}^{(j),n_j=0}(z^j,\bar{z}^j)\equiv f^{I_{AB}^{(j)}}=\left\{\begin{array}{ll}{\Theta}^{I_{AB}^{(j)},M_{AB}^{(j)}}\left(Z_{AB}^j(t)\right) & (M_{AB}^{(j)}>0)\\ (\mathcal A_j)^{-\frac12}& (M_{AB}^{(j)}=0)\\ 0&(M_{AB}^{(j)}<0)\end{array}\right.
\end{align}
where $Z_{AB}^j(t)\equiv z^j+\frac{\zeta_{AB}^{j}(t)}{|M_{AB}^{(j)}|}$, ${\cal A}_j\equiv (2\pi R_i)^2{\rm Im}\tau_j$ is the area of the $j$-th torus and $I_{AB}^{(j)}$ denotes the label of the degeneracy
\begin{align}
    I_{AB}^{(j)}=\left\{\begin{array}{cc} 1,\cdots,\left|M_{AB}^{(j)}\right| &(M_{ab}^{(j)}>0)\\
    0&(M_{ab}^{(j)}>0)\\
    \text{no sol.} &(M_{ab}^{(j)}<0)\end{array}\right. .
\end{align}
We have also introduced
\begin{align}
    \Theta^{I,M}(z)\equiv \mathcal{N}_M e^{\pi\ri\frac{{\rm Im}z}{{\rm Im}\tau}Mz}\vartheta\left[\begin{array}{c}I/M\\ 0\end{array}\right](Mz,M\tau),
\end{align}
where the Jacobi theta function $\vartheta\left[\begin{array}{c}a\\ b \end{array}\right](\nu,\tau)$ is defined below~\eqref{magzeromode} and the normalization factor $N_{M^{(j)}}=(2{\rm Im}\tau_i|M^{(j)}|/{\cal A}_j^2)^\frac14$ is fixed such that 
\begin{align}
    \int dz^jd\bar{z}^j\sqrt{g_j}\bar{f}^If^J=\delta^{IJ},
\end{align}
either for $I,J\neq 0$ or $I=J=0$. For $i\neq j$ sector, the zero mode wave function becomes
\begin{align}
   f_{i,AB}^{(j),n_j=0}(z^j,\bar{z}^j)\equiv f^{I_{AB}^{(j)}}=\left\{\begin{array}{ll}0&(M_{AB}^{(j)}<0)\\ (\mathcal A_j)^{-\frac12}& (M_{AB}^{(j)}=0)\\ 
\left({\Theta}^{I_{AB}^{(j)},M_{AB}^{(j)}}\left(\tilde{Z}_{AB}^j(t)\right)\right)^*& (M_{AB}^{(j)}>0)\end{array}\right.
\end{align}
where $\tilde{Z}_{AB}^j(t)\equiv z^j-\frac{\zeta_{AB}^{j}(t)}{|M_{AB}^{(j)}|}$. Using these wave functions and performing the integration over compact spaces yield the 4D effective K\"ahler and super-potential of zero modes, respectively, as
\begin{align}
   K=&\sum_{i=1}^3\sum_{\bm I_{AB}}{\rm Tr}\left[\bar{\bm Q}^{\bm I_{AB}}_{\bar i}{\bm Q}^{{\bm I}_{AB}}_i\right],\\
   W=&\sum_{i,j,k}\sum_{A,B,C}\sum_{\bm I_{AB},\bm I_{BC},\bm I_{CA}}y_{\bm I_{AB}\bm I_{BC}\bm I_{CA}}^{ijk}(t){\rm Tr}\left[\bm{Q}_i^{\bm I_{AB}}\bm{Q}_j^{\bm I_{BC}}\bm{Q}_k^{\bm I_{CA}}\right]
\end{align}
where ${\bm Q}^{\bm I_{AB}}_i\equiv (\pi R_i)^{-1}g^{-1}{\bm\phi}^{\bm I_{AB},{\bm n}={\bm 0}}_{i}$, we have omitted the couplings to the gauge superfields of residual gauge symmetries and introduced $\bm I_{AB}\equiv(I^{(1)}_{AB},I^{(2)}_{AB},I^{(3)}_{AB})$ and the trace is taken over gauge indices $a,b$. The yukawa coupling is given by overlap integral of zero mode wave functions and acquires time dependence through Wilson lines,
\begin{align}
    y_{\bm I_{AB}\bm I_{BC}\bm I_{CA}}^{ijk}(t)\equiv& -\frac{g\epsilon^{ijk}}{3\sqrt2 }\prod_{r=1}^3\lambda_{ABC}^{(r)}.
\end{align}
Here we have defined 
\begin{align}
    \lambda_{ABC}^{(r)}=&\int d^2z_{r}\sqrt{g_{r}}f^{I_{AB}^{(r)}}f^{I_{BC}^{(r)}}f^{I_{CA}^{(r)}}\nonumber\\
    =&\left\{\begin{array}{cc}\tilde{\lambda}^{(r)}_{AB,C}&(M_{AB}^{(r)}>0)\\ \tilde{\lambda}^{(r)}_{BC,A}&(M_{BC}^{(r)}>0)\\
    \tilde{\lambda}^{(r)}_{CA,B}&(M_{CA}^{(r)}>0)\end{array}\right.,
\end{align}
for the case $M_{AB}^{(r)}M_{BC}^{(r)}M_{CA}^{(r)}>0$ where 
\begin{align}
\tilde{\lambda}_{AB,C}^{(r)}=&{\cal N}^{-1}_{M_{AB}^{(r)}} {\cal N}_{M_{BC}^{(r)}}{\cal N}_{M_{CA}^{(r)}}\sum_{m=1}^{M_{AB}^{(r)}}\delta_{I_{BC}^{(r)}+I_{CA}^{(r)}-mM_{BC}^{(r)},I_{AB}^{(r)}}\nonumber\\
&\times \exp\left[\frac{\pi\ri}{{\rm Im}\tau_r}\left(\frac{\bar\zeta_{AB}^{r}(t)}{M_{AB}^{(r)}}{\rm Im}\zeta_{AB}^{r}(t)+\frac{\bar\zeta_{BC}^{r}(t)}{M_{BC}^{(r)}}{\rm Im}\zeta_{BC}^{r}(t)+\frac{\bar\zeta_{CA}^{r}(t)}{M_{CA}^{(r)}}{\rm Im}\zeta_{CA}^{r}(t)\right)\right]\nonumber\\
&\times\vartheta\left[\begin{array}{c}\frac{M_{BC}^{(r)}I^{(r)}_{CA}-M_{CA}^{(r)}I^{(r)}_{BC}+mM_{BC}^{(r)}M_{CA}^{(r)}}{M_{AB}^{(r)}M_{BC}^{(r)}M_{CA}^{(r)}}\\0\end{array}\right]\left(M_{BC}^{(r)}\bar{\zeta}^{r}_{CA}(t)-M_{CA}^{(r)}\bar{\zeta}^{r}_{BC}(t),-\bar\tau_rM_{AB}^{(r)}M_{BC}^{(r)}M_{CA}^{(r)}\right),
\end{align}
which follows from the fusion rules of Jacobi theta functions. For $M_{AB}^{(r)}M_{BC}^{(r)}M_{CA}^{(r)}=0$, we have $\lambda_{ABC}^{(r)}=\mathcal{A}^{-\frac12}_r$ since one of the modes acquires a constant wave function $\mathcal{A}^{-\frac12}_r$ and the rest yields $1$ from normalization of the mode function. Notice that for $M_{AB}^{(r)}M_{BC}^{(r)}M_{CA}^{(r)}<0$, $M_{AC}^{(r)}M_{CB}^{(r)}M_{BA}^{(r)}>0$, and appropriate relabeling in the above result allows one to use the same formula. In deriving the above formula, one needs a fusion formula for Jacobi theta functions and to take into account the fact that $\zeta^{r}_{AB}+\zeta^{r}_{BC}+\zeta^{r}_{CA}=0$ and $M^{(r)}_{AB}+M^{(r)}_{BC}+M^{(r)}_{CA}=0$. See \cite{Cremades:2004wa} for details.

Let us discuss how to construct semi-realistic models that contain the standard model gauge theories. Following \cite{Abe:2012fj}, we consider $U(8)$ gauge theory with Abelian magnetic fluxes,
\begin{align}
    F_{2+2r,2+3r}=2\pi\left(\begin{array}{ccc}
    M_{C}^{(r)}{\bm 1}_{4\times4} & &\\
    &M_{L}^{(r)}{\bm 1}_{2\times2}&\\
    & & M_{R}^{(r)}{\bm 1}_{2\times2}\end{array}\right),
\end{align}
with which $U(8)\to U(4)\times U(2)\times U(2)$ and we introduce Wilson lines
\begin{align}
    {\bm\zeta}^{r}=\left(\begin{array}{ccccc}
    \zeta^{r}_C{\bm 1}_{3\times3}& & & & \\
    &\zeta^{r}_{C'}& & & \\
     & & \zeta^{r}_L{\bm 1}_{2\times2}& & \\
      & & &\zeta^{r}_{R'}& \\
        & & & &\zeta^{r}_{R''}
    \end{array}\right)
\end{align}
which lead to $U(4)\times U(2)\times U(2)\to SU(3)_C\times SU(2)_L\times (U(1))^5$ and a linear combination of multiple $U(1)$ is identified with $U(1)_Y$. Quantized magnetic fluxes are chosen as
\begin{align}
 {\rm on} \ {\mathbb T}_{1}^2: (M_C^{(1)},M_L^{(1)},M_R^{(1)})=&(0,3,-3)\nonumber\\
 {\rm on} \ {\mathbb T}_{2}^2: (M_C^{(2)},M_L^{(2)},M_R^{(2)})=&(0,-1,0)\nonumber\\
 {\rm on} \ {\mathbb T}_{3}^2: (M_C^{(3)},M_L^{(3)},M_R^{(3)})=&(0,0,1),\label{flux}
\end{align}
and assume the area ratio $\mathcal{A}_{(1)}/{\cal A}_{(2)}=\mathcal{A}_{(1)}/{\cal A}_{(3)}=3$ with which the D-term tadpole terms cancel. The zero modes of chiral superfields that describe matters are
\begin{align}
   & {\bm Q}_1^{\bm I_{AB}}=\left(\begin{array}{cc|c|cc}
    \Omega_{C}^{(1)}&\Xi^{(1)}_{CC'}&0&\Xi_{CR'}^{(1)}&\Xi_{CR''}^{(1)}\\
    \Xi_{C'C}^{(1)}&\Omega_{C'}^{(1)}&0&\Xi_{C'R'}^{(1)}&\Xi_{C'R''}^{(1)}\\ \hline
    \Xi_{LC}^{(1)}&\Xi_{LC'}^{(1)}&\Omega_L^{(1)}&H_u^K&H_d^K\\ \hline
    0&0&0&\Omega_{R'}^{(1)}&\Xi_{R'R''}^{(1)}\\
0&0&0&\Xi_{R''R'}^{(1)}&\Omega_{R''}^{(1)}
    \end{array}\right),\\
    &{\bm Q}_2^{\bm I_{AB}}=\left(\begin{array}{cc|c|cc}
    \Omega_{C}^{(2)}&\Xi^{(2)}_{CC'}&Q^I&0&0\\
\Xi_{C'C}^{(2)}&\Omega_{C'}^{(2)}&L^I&0&0\\ \hline 0&0&\Omega_L^{(2)}&0&0\\ \hline
0&0&0&\Omega_{R'}^{(2)}&\Xi_{R'R''}^{(2)}\\
0&0&0&\Xi_{R''R'}^{(2)}&\Omega_{R''}^{(2)}
    \end{array}\right),\\
    & {\bm Q}_3^{\bm I_{AB}}=\left(\begin{array}{cc|c|cc}
    \Omega_{C}^{(3)}&\Xi^{(3)}_{CC'}&0&0&0\\
\Xi_{C'C}^{(3)}&\Omega_{C'}^{(3)}&0&0&0\\ \hline 0&0&\Omega_L^{(3)}&0&0\\ \hline
U^J&N^J&0&\Omega_{R'}^{(3)}&\Xi_{R'R''}^{(3)}\\
D^J&E^J&0&\Xi_{R''R'}^{(3)}&\Omega_{R''}^{(3)}
    \end{array}\right),
\end{align}
where the column and row correspond to $A,B=1,\cdots,5$. The supersymmetric standard model sector consists of $(H_u^K,H_d^K,Q^I,L^I,U^J,N^J,D^J,E^J)$ whereas the rest $\Omega,\Xi$ denote open string moduli and charged exotic modes respectively. Although we will omit the details, we note that some of moduli can be projected out by orbifold projection on the second and third tori as shown in \cite{Abe:2012fj}. Nevertheless, moduli associated with the first torus $\Omega^{(1)}_A$ $A=C,C',L,R',R''$ are left. In general, it is rather difficult to remove all moduli fields while realizing realistic particle spectrum and flavor structure.\footnote{Exceptional possibility is e.g. $Dp$ ($p<9$) brane system where some of the moduli that appear in 10D Yang-Mills model are absent from the beginning.} Notice that the electric gauge potential on tori correspond to the homogeneous modes of open string moduli, e.g. $\langle\Omega_C^{(1)}(t)\rangle\propto\bar{\zeta}^{1}_C(t)$. We will implicitly impose the orbifold projection on the 2nd and 3rd torus with which Yukawa couplings appear only for the standard model sector.

The magnetic fluxes are chosen such that quark and lepton multiplets $(Q,U,D,L,N,E)$ have three zero modes namely three generations. Indeed, the set of flux differences are
\begin{align}
    Q,L:&\  M_C^{(1)}-M_L^{(1)}=-3,\ M_C^{(2)}-M_L^{(2)}=+1,\ M_C^{(3)}-M_L^{(3)}=0,\\
    H_u,H_d:&\ M_L^{(1)}-M_R^{(1)}=+6,\ M_L^{(2)}-M_R^{(2)}=-1,\ M_L^{(3)}-M_R^{(3)}=-1,\\
    U,D,N,E:&\ M_R^{(1)}-M_C^{(1)}=-3,\ M_R^{(2)}-M_C^{(2)}=0,\quad M_R^{(3)}-M_C^{(3)}=+1.
\end{align}
Notice that in such a case, six generations of Higgs sector $(H_u,H_d)$ unavoidably appear, which is important to realize flavor structure such as particle mass spectrum and Cabibbo-Kobayashi-Maskawa (CKM) matrix consistently with experimentally observed values. The VEVs of Higgs fields are determined by $\mu$-terms that cannot appear at the perturbative level but may show up at the orbifold fixed point where the bulk supersymmetry is broken to at most 4D $\mathcal{N}=1$. Such $\mu$-terms are generated through non-perturbative effects. In the following, we simply assume Higgs VEVs as an illustration.

Wilson lines are also responsible for flavor structure and we rename the Wilson lines on the first torus as
\begin{align}
    &\zeta_Q(t)\equiv \zeta^1_C(t)-\zeta^1_L(t),\quad \zeta_U(t)\equiv\zeta^1_{R'}(t)-\zeta^1_{C}(t),\quad \zeta_D(t)\equiv\zeta^1_{R''}(t)-\zeta^1_C(t),\nonumber\\
   & \zeta_L(t)\equiv \zeta^1_{C'}(t)-\zeta^1_L(t),\quad \zeta_N(t)\equiv\zeta^1_{R'}(t)-\zeta^1_{C'}(t),\quad \zeta_E(t)\equiv\zeta^1_{R''}(t)-\zeta^1_{C'}(t).
\end{align}
We also define
\begin{align}
    \zeta_{H_u}\equiv \zeta^1_L(t)-\zeta^1_{R'}(t),\quad \zeta_{H_d}\equiv \zeta_L^1(t)-\zeta_{R''}^1(t),
\end{align}
which are not independent of the above Wilson lines, for instance $\zeta_Q+\zeta_U+\zeta_{H_u}=0$.

The Yukawa couplings are essentially determined only by the fluxes and Wilson lines on the first torus only: Notice that on the 2nd and 3rd torus, respectively, there appears a superfield having zero flux difference $M_{AB}=0$ and therefore, the overlapping integral becomes trivial. Thus, only wave functions on the 1st torus yield nontrivial flavor structure. Notice that realistic flavor structure is realized only by including Wilson lines such that e.g. the left-handed quark and lepton obtain different wave function profiles. The continuous Wilson lines are allowed only on the 1st torus, and we identify them to be the electric gauge potential.\footnote{The orbifold conditions only allow discrete values of Wilson lines and there appear no moduli for such sector as expected. We here assume no discrete Wilson lines on the 2nd and 3rd torus.}  
With the three generation Ansatz for magnetic fluxes~\eqref{flux}, the Yukawa superpotential is 
\begin{align}
   W=& y^{({\cal Q}_R)}_{IJK}\Phi_{{\cal Q}_L}^I\Phi_{{\cal Q}_R}^J\Phi_{{\cal Q}_H}^K\nonumber\\
   =&y^{(U)}_{IJK}Q^IU^JH_u^K+y^{(D)}_{IJK}Q^ID^JH_d^K+y^{(N)}_{IJK}L^IN^JH_u^K+y^{(E)}_{IJK}L^IE^JH_d^K,\label{Yukawaterms}
\end{align}
where we have defined labels $\{{\cal Q}_L\}=\{Q,L\}$, $\{{\cal Q}_R\}=\{U,D,N,E\}$ and $\{{\cal Q}_H\}=\{H_u,H_d\}$. Then, the Yukawa couplings are given by
\begin{align}
y_{IJK}^{({\cal Q}_R)}=&\frac{g(3{\rm Im}\tau_1)^\frac14}{\sqrt{2\mathcal{A}_1\mathcal{A}_2\mathcal{A}_3}}\exp\left[\frac{\pi\ri}{{\rm Im}\tau_1}\left(\frac{\bar\zeta_{{\cal Q}_H}(t)}{6}{\rm Im}\zeta_{{\cal Q}_H}(t)-\frac{\bar\zeta_{{\cal Q}_L}(t)}{3}{\rm Im}\zeta_{{\cal Q}_L}(t)-\frac{\bar\zeta_{{\cal Q}_R}(t)}{3}{\rm Im}\zeta_{{\cal Q}_R}(t)\right)\right]\nonumber\\
&\times \sum_{m=1}^{6}\delta_{I+J+3(m-1),K}\vartheta\left[\begin{array}{c}\frac{3(I-J)+9(m-1)}{54}\\0\end{array}\right]\left(3\bar{\zeta}_{{\cal Q}_L}(t)-3\bar{\zeta}_{{\cal Q}_R}(t),-54\bar\tau_1\right),\label{Yukawa}
\end{align}
with the understanding that the set of $({\cal Q}_L,{\cal Q}_R,{\cal Q}_H)$ appearing above has to be the gauge invariant ones shown in the second line of \eqref{Yukawaterms}. Note that the Wilson lines identically satisfy $\zeta_{{\cal Q}_L}(t)+\zeta_{{\cal Q}_R}(t)+\zeta_{{\cal Q}_H}(t)=0$ by definition.

\bibliographystyle{JHEP}
\bibliography{main.bib}
\end{document}